\documentclass[10pt,conference,letterpaper]{IEEEtran}
\usepackage{romannum}
\usepackage{cite}
\usepackage[pdftex]{graphicx}
\usepackage{algpseudocode}
\usepackage{algorithm}
\usepackage{bm}
\usepackage{multirow,url,float,epstopdf}
\usepackage{enumerate}
\usepackage[cmex10]{amsmath}
\usepackage{tikz} 
\usepackage{amssymb}
\usepackage{bbm}
\usepackage{verbatim}
\usepackage{bbm}
\newtheorem{lemma}{Lemma}
\newtheorem{theorem}{Theorem}

\newtheorem{remark}{Remark}

\newtheorem{assumption}{Assumption}

\ifCLASSINFOpdf

\else

\fi

\begin{document}
\title{ Throughput Optimal Decentralized Scheduling
of Multi-Hop Networks with End-to-End
Deadline Constraints: II\\ Wireless Networks with Interference}
\author{Rahul~Singh, P.R. Kumar, and Eytan Modiano
\thanks{Rahul Singh and Eytan Modiano are with the Laboratory of Information and Decision Systems (LIDS), Massachusetts Institute of Technology, Cambridge, MA 02139, USA.
        {\tt\small rsingh12@mit.edu, modiano@mit.edu}}%
\thanks{P. R. Kumar is at Dept. of ECE, Texas A\&M Univ., 3259 TAMU, College Station, TX 77843-3259
		{\tt\small prk@tamu.edu}}
		}
\maketitle
\begin{abstract}
Consider a multihop wireless network serving multiple flows in which wireless link interference constraints are described by a link interference graph. For such a network, we design routing-scheduling policies that maximize the end-to-end timely throughput of the network. Timely throughput of a flow $f$ is defined as the average rate at which packets of flow $f$ reach their destination node $d_f$ within their deadline. 

Our policy has several surprising characteristics. Firstly, we show that the optimal routing-scheduling decision for an individual packet that is present at a wireless node $i\in V$
is solely a function of its location, and ``age". Thus, a wireless node $i$ does not require the knowledge of the ``global" network state in order to maximize the timely throughput. We notice that in comparison, under the backpressure routing policy, a node $i$ requires only the knowledge of its neighbours queue lengths in order to guarantee maximal stability, and hence is decentralized. The key difference arises due to the fact that in our set-up the packets loose their utility once their ``age" has crossed their deadline, thus making the task of optimizing timely throughput much more challenging than that of ensuring network stability. Of course, due to this key difference, the decision process involved in maximizing the timely throughput is also much more complex than that involved in ensuring network-wide queue stabilization. In view of this, our results are somewhat surprising. 

Secondly, the complexity of algorithms that obtaining the policy scales linearly with the number of links present in the network. In case the network parameters are unknown, we derive iterative ``online learning" algorithms that yield the optimal policy.

We divide the available bandwidth into multiple sub-channels, and allow a policy to activate a set of non-interfering links on each of the subchannels. We consider several types of constraints on the bandwidth availability. When the wireless network is constrained by the average bandwidth utilization, then the obtained policy is shown to be optimal. 

In case the wireless network has to operate under a hard constraint on the available bandwidth, we truncate the policy obtained for the network with average bandwidth constraints in order that the bandwidth utilized by it is $\leq K$ units in each time-slot. We show that this truncated policy is asymptotically optimal as the network traffic is scaled to $\infty$. 
\end{abstract}

\section{Introduction}\label{sec:intro}
For multi-hop networks serving real-time applications, data packets typically have a deadline and it is important to ensure that a maximum fraction of the packets reach their destination within the deadline. Currently the performance metric of throughput optimality, i..e, the property of ensuring queue stability for maximal set of arrival vectors, is widely popular in designing network control policies. Backpressure policy is known to be throughput optimal under very general conditions~\cite{tassi1,neelyinfoc}. Similarly the Q-CSMA scheme, which combines backpressure routing along with the CSMA algorithm to find the maximum weight matching, is known to be throughput optimal for scheduling traffic in wireless networks under link interference constraints~\cite{jiang2010distributed,srikant1}. However, the goal of a throughput maximizing policy does not amount to ensuring that the packets meet stringent end-to-end deadlines. Thus, for example, the backpressure policy, or its wireless version Q-CSMA, is known to have poor performance with respect to average delays~\cite{d1,d2,d3}. Thus, we consider the problem of scheduling packets for multi-hop wireless interference network in which data packets have stringent deadlines that has to be met.  

However, designing an optimal policy for maximizing the timely throughput is much more complicated than ensuring the throughout optimality since the timely throughput attained by a policy is ``highly sensitive" to its routing-scheduling decisons. This is the case because the utility earned from a single packet ``drops to $0$" in a discontinuous fashion as soon as its age has crossed its deadline, and hence even a minute fluctuation in the network bandwidth or a small deviation from the optimal decision affects the timely thoughput significantly.

\begin{figure}[h]
	\centering
	\includegraphics[width=0.5\textwidth]{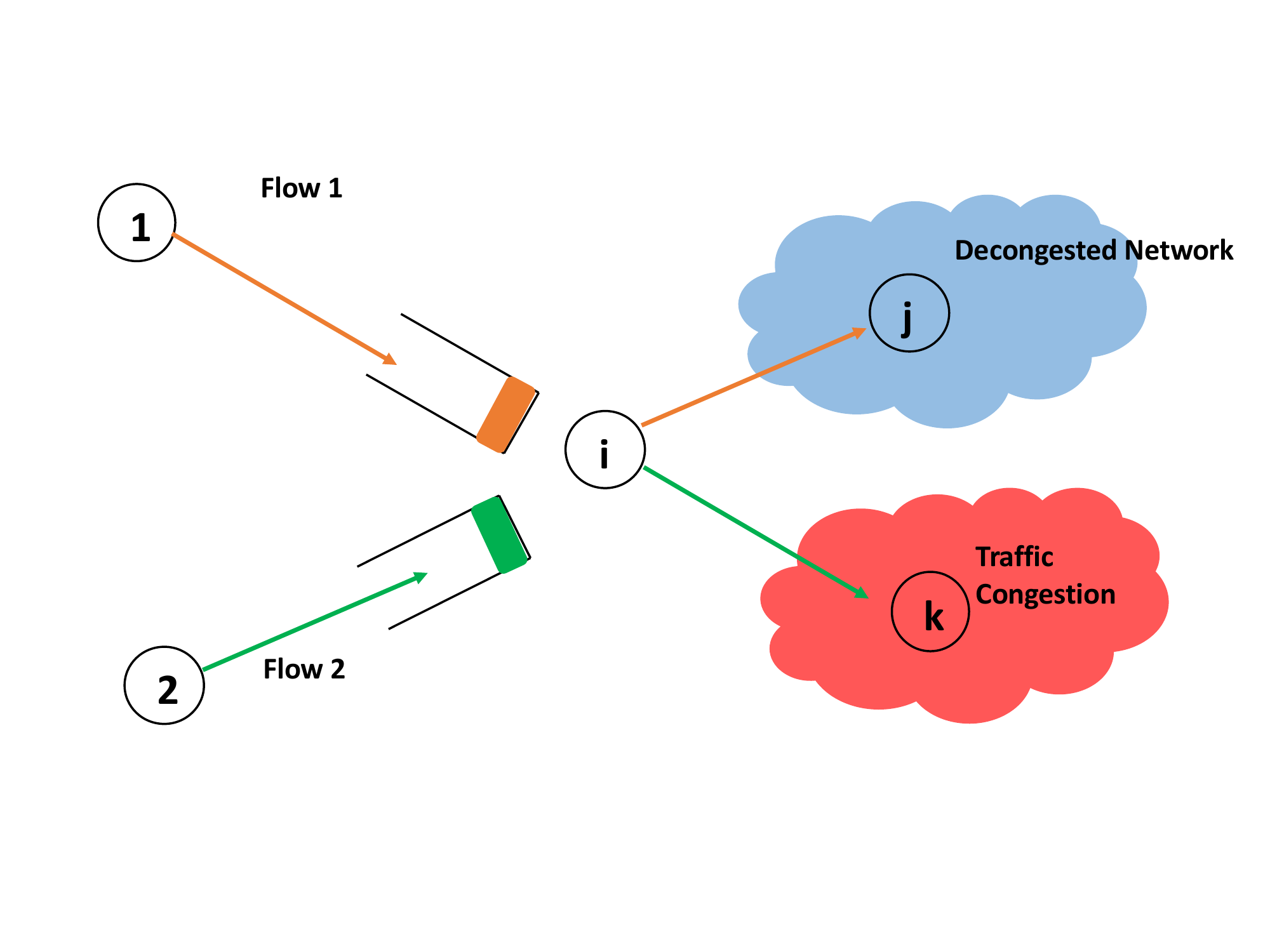}
	\caption{Making optimal scheduling decisions for meeting deadline constraints is a challenging problem that requires the knowledge of the network state. Consider the decision process involved in routing packets at a wireless node $i$. The link pairs $(1,i),(i,j)$ and $(2,i),(i,k)$ interefere with each other. Thus, either the pair $(1,i),(i,k)$ or $(2,i),(i,j)$ can be activated simultaneously. Since Flow 2's traffic faces downstream congestion, it might be ``optimal" to exclusively focus on scheduling Flow $1$'s packets. This is true because a packet sent on link $(i,k)$ may have to wait for long period due to traffic congestion, and hence will not be able to make it to its destination node within deadline. Thus the links $(1,i)$ and $(i,j)$ should be given priority. However, since $(1,i)$ and $(i,j)$ interfere, one further has to make a choice between activating one link amongst these two. It may be the case that the packets that can be scheduled on $(i,j)$ are ``nearing their deadline", and hence link $(i,j)$ should be prioritized over $(1,i)$. Alternatively, it may be the case that these packets have little chances of making it to the destination, because their ``deadline has almost crossed", and $(i,j)$ has a low channel reliability. In this case, link $(1,i)$ must be activated. We thus notice that the decisions have to be made on the basis of the state of the network. Thus, a knowledge of the complete network state is required at each time $t$ in order to maximize the timely throughput. The presence of wireless interference makes the problem further difficult since a choice of an independent set (whose number is exponential in $|E|$) of links has to be made.}
	\label{fig:wireless-1}
\end{figure}

For a network control policy, it is highly desirable that it is decentralized, meaning that the wireless nodes do not need to know the ``global" state of the system denoted $X(t)$. The backpressure policy is decentralized, in the sense that a node needs to know the queue lengths of only its neighbouring nodes, i.e, the nodes that are connected to it via an outgoing link. However, as depicted in Fig.~\ref{fig:wireless-1}, it seems unlikely that the timely throughput maximization problem has a decentralized solution. 

In this paper we derive a new class of policies that not only maximize the timely throughput, but are also \emph{highly decentralized} in the sense that a wireless ndoe only needs to know the age of the packets present with it in order to make scheduling decisions. This eliminates the need to share any information amongst the nodes. 

In a companion paper~\cite{singh2016throughput}, we developed a theoretical framework and proposed decentralized policies that maximize the timely-throughput in a multi-hop stochastic network. This paper extends the ideas therein to the set-up of wireless networks in which the links suffer from wireless interference. 
\section{Contributions and Past Works}
We consider the problem of designing efficient scheduling policy for multihop wireless networks in which the data packets have a deadline associated with their deliveries. 

 As described in Fig.~\ref{fig:wireless-1}, the timely throughput is ``highly sensitive" to the routing-scheduling decisons because the utility earned from a single packet drops to $0$ in a discontinuous fashion as soon as its age crosses its deadline. Thus, we cannot use the fluid model, which is commonly used in combination with Lyapunov techniques in order to establish throughput optimality. We have to resort to directly solving the stochastic network model rather than its fluid approximation. The policy provided by us is decentralized, and its complexity scales linearly with the network size, thus addressing the crucial problem of meeting end-to-end packet deadlines, that is typically encountered in multi-hop scheduling in wireless networks.
 

We pose the problem of finding the timely throughput maximizing policy as an Markov Decision Process (MDP) in Section~\ref{sec:tpm}. However, the MDP formulation is intractible since the resulting policy is centralized. We then devise a novel approach to decentralized stochastic control under constraints in Section~\ref{sec:rp} by replacing the hard versions of the i) link interference, and ii) network-wide bandwidth availability constraints, by their ``softer versions" which involve time averages of the corresponding constraint violations. This relaxation is in inspired by Whittle's relaxation for the Restless Multiarmed Bandit problem. This relaxation of the constraints to their soft versions leads us to a constrained MDP (CMDP).

We first deal exclusively with the optimal scheduling problem under link-level average constraints, and hence ignore the wireless interference completely. Since the primal CMDP is intractible we then consider the dual version of the CMDP, and show that it is much simpler to solve. This reduction in complexity results from the realization that the Lagrangian with the multiplier set equal to $\lambda=\{\lambda_{\ell}\}$
can be viewed as the sum of the rewards earned by the individual data packets, where the reward of a single packet is equal to its timely throughput minus the price it pays for using network bandwidth. Since the optimization of reward of an individual packet can be performed independently of other packets present in the network, this means that the evaluation of dual function $D(\lambda)$, and consequently its optimization ($\max_{\lambda\geq 0}D(\lambda)$) can be performed in a decentralized manner. We then use the strong duality property of linear programs  
to deduce a highly decentralized policy denoted $\pi^\star$ for the primal CMDP that can be computed in a distributed way. 

We then come back to original problem of scheduling under wireless interference constraints. We introduce a new variant of the commonly used CSMA protocol in order to allocate the total bandwidth amongst various independent sets in a decentralized way. Then, we consider a scaling of the network, under which the total available bandwidth, and the traffic arrival rates are scaled by a parameter $N$. The total bandwidth is then divided into $N$ ``orthogonal" channels, and CSMA protocol is used on each of them independently. The policy $\pi^\star$, that was optimal under link-level average bandwidth constraints is then modified to yield a policy $\tilde{\pi}$ for the network with interference. We then use show that as $N\to\infty$, the timely throughput of $\tilde{\pi}$ approaches within $O(\sqrt{N})$ of the optimal policy, and hence $\tilde{\pi}$ is asymptotically optimal. The proof relies on the structure of the policy $\tilde{\pi}$, combined with a ``bandwidth smoothing" that is achieved by using independent CSMA counters on each of the orthogonal subchannels.

We highlight some key differences between the manner in which we utilize the CSMA protocol, from the commonly used CSMA. CSMA has been commonly used to provide decentralized channel access. The CSMA ``aggression parameter" can be modulated, as in~\cite{wlrnd1} based on the network queue lengths in order to ensure the throughput optimality property. Since the focus of these works has been on throughput optimality, by letting the aggression rate $r_\ell$ of the CSMA counters continually adapt to the mismatch between the traffic arrival intensity at link $\ell$ and the time-average bandwidth provided to the link $\ell$ (which is an increasing function of $r_\ell$), the Q-CSMA scheme achieves its throughput optimality. However, since in our case, the contribution of a single packet towards the timely throughput heavily depends on the actual value of the channel capacities on each link $\ell$ belonging to its source-destination path, we need to ensure that the bandwidth fluctuations of each link $\ell$ in the network are ``small" over the entire duration that a packet remains in the network. Thus, we divide the total available bandwidth into $N$ sub-channels, each of unit bandwidth, and assume that each link $\ell$ utilizes independent CSMA counters on each of these $N$ orthogonal sub-channels in order to attain channel access. Due to ``central limit theorem type" bandwiwdth smoothing, and due to the property of $\tilde{\pi}$, we are able to ensure that the network timely throughput is not affected much by having CSMA based channel access mechanism, as opposed to a centralized controller. This optimality result can be viewed as a large deviations control of timely throughput of wireless networks.

The CSMA aggression parameter $r$ is adjusted based on the feedback received on the link prices, and we utilize a gradient decent based scheme in order to converge to a locally optimal solution. Such a deficiency arises because the optimization problem is non-convex.

The problem of scheduling traffic in order to satisfy timely throughput constraints of multiple clients being served by a single-hop wireless network is fairly well-understood by now~\cite{hbk,IHHou10Utility,hou2012real,rahul1,rahul,IHHou10Hoc,hou2011broadcasting,hou2011survey,houkum13,hou2011optimality}. For this single-hop network, simple to implement greedy policies are shown to be optimal. However, the problem of extending the approach to a multi-hop network has been open for quite some time, though several heuristics have been proposed~\cite{atilla3}, or a fraction of the maximum achievable timely-throughput has been attained~\cite{mao2016optimal}. 

\section{System Description}\label{sec:sd}
\emph{Network Description} : The wireless network is described by a graph $G=(V,\cal{E})$, where $V$ is the set of wireless nodes, and $\cal{E}$ is the set of directed edges of the form $(i,j),~i,j\in V$. Associated with each edge $\ell=(i,j)$ is a
rate $C_\ell$ denoting that node $i$ can attempt transmission of $C_\ell$ packets to node $j$ in a single time slot. A packet transmission can fail due to to either the unreliability of the wireless channel,
or the interference caused by other concurrent wireless transmissions. 
Assuming that no interfering links are transmitting, a transmission on a link $\ell$ in a time slot $t$ succeeds with a probability $p_\ell$. The quantity $p_\ell$ is called the channel reliabilty of link $\ell$. Throughout we assume that the random outcomes of packet transmissions are independent across time and links. 

\emph{Interference Model :} We model wireless interference constraints via an ``edge interference graph". It is an undirected graph in which each vertex corresponds to a wireless link from the set $\mathcal{E}$. There is an undirected edge between two nodes $\ell_1,\ell_2$ if and only if the links $\ell_1$ and $\ell_2$ interfere, i.e., if $\ell_1,\ell_2$ are activated simultaneously, then the packet transmissions occurring on both the links fail.

Let $\cal{I}$ be the set comprising of the~\textit{maximal independent sets} of the link interference graph. We denote the maximal independent sets by $IS_1,IS_2,\ldots$. Each maximal independent set $IS \in \cal{I}$ consists of links $\ell \in \mathcal{E}$ that do not interfere with each other, because they are not connected by an edge in the edge interference graph.
Moreover any link $\ell$ that does not belong to $IS$ definitely interferes with at least one link in $IS$ because $IS$ is maximally independent.

Therefore, at any time $t$, a scheduling policy $\pi$ is restricted to choosing an $IS\in \mathcal{I}$ and activating the corresponding links. 

\emph{Multiple Flows:} The network is shared by $F$ flows, where each flow $f$ has a source node $s_f$ and a destination node $d_f$. We suppose that time is slotted, and the network evolves at tim slots $t=1,2,\ldots$. The time-duration of a single time-slot is equal to the time taken to attempt transmission of $C_\ell$ packets on link $\ell\in \mathcal{E}$. 
The packets for flow $f$ arrive at their source $s_f$ in an i.i.d. fashion across time. We assume that the packet arrival process is uniformly bounded across flows and time. Since a source-destination pair $(s_f,d_f)$ need not be a link, packets will typically need to traverse multihop paths before reaching their destination.
\section{Multi-hop Timely Throughput } \label{sec:tpm}
We define the timely throughput metric for a multihop wireless network described in the previous section. Let $\tau \geq 1$ be an integer which represents the deadline for packet deliver, i.e., it is the maximum allowable end-to-end delay
that a packet may incur.

\emph{ Multihop Timely Throughput:} We define the timely throughput $\bar{D}_f $
of a flow $f$ as the average number of packets per unit time of flow $f$ that reach the destination node $d_f$ within $\tau$ time-slots from the time they are generated at the source node $s_f$\footnote{Throughout the paper, we assume that the averages corresponding to stochastic processes of interest converge almost surely. This is not restrictive since we are optimizing over finite state Markov processes, which always admit a stationary policy that is optimal. Since a time-homogenous finite-state Markov chain is necessarily positive recurrent, they admit almost sure limits. 
Thus, we can replace $\limsup$ by $\lim$ etc.}, 
\begin{align*}
\bar{D}_f := \lim_{T\to\infty}\frac{1}{T}\mathbb{E}\left\{\sum_{t=1}^{T}D_f(t)\right\},
\end{align*}
where $D_f(t)$ is the number of flow $f$ packets delivered at time $t$ to the destination node $d_f$ within $\tau$ time-slots from the time they were generated at $s_f$. The expectation is under the probability measure induced by the scheduling policy $\pi$ under use, the packet arrival processes and the random states of wireless links. 
We will denote the average value of $D_f(t)$ by $\bar{D}_f$, so that $\bar{D}_f$ is the timely throughput of flow $f$. Similarly, time-averages corresponding to a stochastic process $X(t)$ will be denoted by $\bar{X}$.

\emph{ Network State:} 
The state of an individual packet at time $t$ is described by,
\begin{enumerate}
\item the node $i\in V$ at which it is present, and,
\item the age (time elapsed since generation) of the packet.\footnote{ We note that the age of a packet can be easily deduced by time-stamping it at the time it was generated at its source node.}
\end{enumerate}
The state of the network at time $t$, denoted $X(t)$, is described by specifying
the state of each packet that is present in the network, for each flow $f$.
We note that for conventional networks that are designed to be throughput or delay optimal do not need to keep track of the ages of packets. However, in our case the age of a packet has to be taken into account because a packet delivered after its age has crossed $\tau$ does not contribute to timely throughput.

\emph{ Policy:} A policy $\pi$ maps $X(t)$, the system state at time $t$, to an action $U(t)$, which is uniquely described by the following i) the independent set $IS$ that has to be activated at time $t$, ii) the packets that are to be transmitted on each link $\ell\in IS$. Thus, a policy $\pi$ has to make both routing and scheduling decisions simultaneously.   

\emph{ The MDP for Timely Throughput Maximization:} The problem of maximizing a weighted sum of timely throughputs can be posed as the following MDP,
\begin{align}\label{mdp}
\max_\pi \sum_f \beta_f \bar{D}_f,
\end{align}
where $\pi$ is a policy, and the weights $\beta_f\geq 0$ allow the network operator to prioritize the various packets on the basis of their relative importance of their timely-throughputs.

\emph{Complexity of Solving the MDP~\eqref{mdp}:} The size of the state-space corresponding to MDP~\eqref{mdp} is $\left(|V|\tau\right)^B$, where $B$ is a bound on the number of packets that can be present in the network at any given time. Thus it is computationally infeasible to solve~\eqref{mdp}. Secondly, the resulting optimal policy will prescribe the optimal action $U(t)$ as a function of the network state $X(t)$, and hence calls for a centralized controller. This again is quite impractical.

\section{Orthogonal Channels}\label{sec:oc}
Our results will be asymptotic in nature, i.e., the resulting policy will be shown to be nearly optimal as the network capacity is scaled to $\infty$. Therefore, we will slightly enhance the system model. 
  
We will assume that the total bandwidth available to the wireless network is equal to $K$ units,
divided into $K$ orthogonal sub-channels, so that each orthogonal channel has access to a unit amount of bandwidth. At each time $t$, a scheduling policy $\pi$ can activate an independent set of links on each of the $K$ orthogonal sub-channels.

Denote by $I_m(t)\geq 0$ the amount of bandwidth provided to independent set $IS_m$ at time $t$,
with $\sum_{m}I_m(t)=K$. We suppose that the data rate attainable per
unit bandwidth on link $\ell$ is $C_\ell$ packets/slot. 
Hence the number of packet transmissions that can be scheduled on a link $\ell$ at time $t$ is given by $C_\ell\sum_{m:\ell\in IS_m}I_m(t)$.

In the below, we let $I^\pi(t)$ denote the vector with entries $I^\pi_m(t)$, while $\bar{I}^\pi$ contains their time-averages. Let $\pi$ be any history dependent policy that decides the following at each time $t$: 
i ) $I^\pi_m(t)$, the 
i) number of sub-channels allocated to independent set $IS_m$, and ii) the packets scheduled on each link $\ell\in \mathcal{E}$. 
The timely throughput maximization problem can now be re-stated as,
\begin{align}\label{op:int1}
&\max_{\pi} \sum_{f}\beta_f \bar{D}_f, \\
&\mbox{ s.t. } \sum_m I^\pi_m(t) \leq K, ~~\forall t=1,2,\ldots.\label{op:int2}
\end{align}
under the Markov Decision Process Model.

%
\section{Relaxing the constraints}\label{sec:rp}
Problem~\eqref{op:int1}-\eqref{op:int2} imposes a ``hard" constraint \eqref{op:int2}
on the bandwidth consumption of a feasible policy $\pi$ in \emph{every time-slot} $t=1,2, \ldots$.
We will now consider a simpler version of the bandwidth constraint, by replacing the constraint $\sum_m I^\pi_m(t) \leq K, ~~\forall t=1,2,\ldots$ by a ``soft" version that requires only that the \emph{time-average bandwidth consumption} by $\pi$ is less than $K$ units. 

We pose this relaxation of the timely throughput maximization problem~\eqref{op:int1} as,
\begin{align}
&\max_{\pi} \sum_{f}\beta_f \bar{D}_f\mbox{, such that }\label{eq:intavg} \\
&\qquad \sum_m\bar{I}^\pi_m \leq K \label{const1}, \end{align}
where $\bar{I}_m$ denotes the time-average of the bandwidth ``consumed" by the independent set $IS_m$. The relaxed constraint now allows a policy $\pi$ to utilize more than $K$ units of bandwidth at any time-slot $t$.

We note that that the relaxed constraint \eqref{const1}
$\sum_m\bar{I}^\pi_m \leq K$
is still a constraint of a ``global spatial nature"~\footnote{This is true because the choice of $I_m(t)$ affects the amount of bandwidth available to each link $\ell\in IS_m$, and hence must necessarily depend on the state of the packets present at each link $\ell\in IS_m$.}. Thus, a
An optimal $\pi$ has to achieve a ``global spatial coordination" amongst the various links $\ell\in \mathcal{E}$, and thus the problem~\eqref{eq:intavg}-\eqref{const1} is challenging. 
Hence we will now further relax the constraints, and this will lead us to a somewhat weaker form of 
the hard interference constraints under which any concurrent transmissions on two interfering links fail.

We observe that under a policy $\pi$, the transmission rate obtained by a link $\ell$ at time $t$ is given by
\begin{align}
C^\pi_\ell(t) := C_\ell\sum_{m:\ell\in IS_m}I_m(t),\label{eq:wic} 
\end{align}
while the time-average transmission rate that it receives is given by
\begin{align}
\bar{C}_{\ell}^\pi := C_{\ell} \left(\sum_{m:\ell\in \mathcal{E}_m} \bar{I}_m \right).
\end{align}
The relation between $C^{\pi}_\ell(t)$ and $I(t)$ can be viewed as interference constraint. Thus,
the ``Relaxed Problem" in which the hard interference
  and the hard bandwidth availability constraints have been relaxed to their average versions is as follows,
\begin{align}
&\max_{\pi} \sum_{f}\beta_f \bar{D}_f\mbox{, such that }\label{eq:intavg1} \\
&~ \bar{C}_{\ell}^\pi \leq C_{\ell} \left(\sum_{m:\ell\in IS_m} \bar{I}_m \right),\label{eqintavg12}\\
&~\sum_{m}\bar{I}_m \leq K.\label{eqintavg13}
\end{align}
We interpret the constraints~\eqref{eqintavg12}-\eqref{eqintavg13} as follows. Once a feasible vector $\bar{I}=\{\bar{I}_m\}$ satisfying $\sum_m\bar{I}_m\leq K$ has been fixed, the average link bandwidths $\bar{C}^{\pi}_\ell$ get fixed according to~\eqref{eqintavg12}. The constraint~\eqref{eqintavg12} does not impose any constraint on instantaneous link bandwidths $C^{\pi}_\ell(t)$. They also do not impose any interference constraints, i.e., the edge interference graph model described in Section~\ref{sec:sd} is no longer valid. Hence two links $\ell,\hat{\ell}$ that are connected in edge interference graph, are allowed to carry out concurrent packet transmissions without packet drops\footnote{Equivalently  we can assume that each link $\ell$ has access to unlimited bandwidth, and that the channel it uses is orthogonal to the channels used by other links. }.    

Notice that though we have rid ourselves of the hard interference constraints ~\eqref{eq:wic}, we have retained the constraints i) the average bandwidth available to a link $\ell$ depends on the average bandwidth allocation vector $\bar{I}$, and, ii) the total average bandwidth available to the wireless network, i.e., the constraint $ \sum_m\bar{I}^\pi_m \leq K$. We have only relaxed the constraints~\eqref{eq:wic} imposed by the ``wireless interference system" since they are overly restrictive. This relaxation, in a certain sense, is equivalent to ``weaking" the wireless interference constraints, i.e., under the relaxed constraints an admissible scheduling policy $\pi$ does not need to assign an independent set $IS$ on each of the $K$ sub-channels. However, as will be shown in this paper, solving this relaxed problem does yield a near-optimal solution to the original problem when the network traffic is scaled to $\infty$, and hence the efforts to solve the relaxed problem will not go in vain.

\emph{Connections with Whittle's Relaxation for the MABP} \\The relaxation that we have introduced above is in the sprit of Whittle's relaxation~\cite{whittle} for Restless Multi-Armed bandit problem (MABP)~\cite{Whittle2011Book} which can be posed as an MDP. In the MABP set-up, a controller has to play one ``arm" at each time-slot $t=1,2,\ldots$, and he obtains a reward which is a function of the state of the arm that he currently played. The various arms of the bandit are essentially controlled Markov processes. Whittle's relaxation for the MABP is to replace the hard constraint that only a single arm be played at each time, by a softer constraint that requires that the player plays a single arm only on an average. Since in the relaxed problem, the ``decision processes" for the arms  are decoupled, the relaxed problem is much easier to solve, and the complexity reduces drastically as compared to the MABP. The Whittle's index policy then activates the arm that attains the ``highest gain in reward" from activating it. See~\cite{Whittle2011Book} for details. In our set-up, the packets that are to be routed over the wireless network are the analogues of bandit arms in the MABP. The player is the policy $\pi$ that has to be designed by the network operator, the constraints are the wireless interference constraints. The wireless interference constraints are obviously much more complex than the constraints imposed in the MABP in that only a single independent set/arm has to be played in a time-slot. However, as we will see, the idea of Whittle's relaxation does turn out to be useful even in this complex setup. 

In view of the above discussion, our approach to solving the original problem~\eqref{op:int1}-\eqref{op:int2} will be as follows. We will first solve the much simpler relaxed problem~\eqref{eq:intavg1}-\eqref{eqintavg13} since it is tractable and admits a neat decentralized solution. Denote the solution to~\eqref{eq:intavg1}-\eqref{eqintavg13} by $\pi^\star$. Once $\pi^\star$ has been obtained, we will modify it appropriately, and combine it with CSMA protocol. This will yield us a policy $\tilde{\pi}$ that is \emph{feasible for the original problem}~\eqref{op:int1}. Then, we will show that $\tilde{\pi}$ is asymptotically optimal in the limit the traffic arrival rates, and available bandwidth $K$ are scaled to $\infty$.  
\begin{remark}
 We note that the relaxed version of the link capacity constraints~\eqref{eqintavg12}-\eqref{eqintavg13} still induces a weaker version of the interference constraint.
In later sections we will show that an optimized version of the CSMA protocol resolves the bandwidth allocation problem in a decentralized fashion. Moreover, since the amount of bandwidth it assigns to each link $\ell$ is nearly a constant, i.e., its stochastic fluctuations are small, we would be interested in developing a scheduling policy under the assumption that the bandwidth for each link $\ell$ has been fixed. The reason why we derive the scheduling policy under fixed average link bandwidths is that we will require the solution of this problem in order to optimize the CSMA protocol.
\end{remark}

We begin by addressing the relaxed problem~\eqref{eq:intavg1}-\eqref{eqintavg13} for the case when the link-level bandwidths have been fixed at $C^{av}$.

\section{Scheduling under Link-Level Average Bandwidth Constraints}\label{sec:llbc}
In this section we will be concerned with maximizing the total timely throughput under the constraint that the average bandwidth provided to each link $\ell$ is less than or equal to $C^{av}_{\ell}$. Equivalently, the constraints~\eqref{eqintavg12}-\eqref{eqintavg13} in the problem~\eqref{eq:intavg1}-\eqref{eqintavg13} will be replaced by the constraints 
$$
\bar{C}_{\ell}^\pi \leq C^{av}_{\ell}, \forall \ell \in\mathcal{E},
$$
where $C^{av}_{\ell}$ is the bound on average bandwidth of link $\ell$. We also let $C^{av}=\{C^{av}_\ell\}_{\ell\in E}$. 


Thus, in this Section we will solve the following CMDP,
\begin{align}
f\left(C^{av}\right):&=\max_{\pi} \sum_{f}\beta_f \bar{D}_f\mbox{, such that }\label{eq:p1} \\
&~ \bar{C}_{\ell}^\pi \leq C^{av}_{\ell},\forall \ell \in E.\label{eq:c1}
\end{align}
We will obtain a computationally simple and decentralized solution. However, as discussed in Section~\ref{sec:tpm}, a naive approach to solve the above CMDP in its primal form using the linear programming approach is impractical owing to the curse of dimensionality, and the requirement of a centralized controller.

In order to develop a decentralized and computationally feasible iterative solution, we consider the dual problem associated with the primal CMDP~\eqref{eq:p1}-\eqref{eq:c1}. For a scheduling policy $\pi$, the Lagrangian corresponding to~\eqref{eq:p1}-\eqref{eq:c1} is given by,
\begin{align*}
&\mathcal{L}(\pi,\mu) \\
&= \sum_{f} \beta_f \bar{D}_f  - \sum_\ell \mu_\ell \left(\bar{C}^\pi_\ell- C^{av}_{\ell}\right)\\
&= \sum_{f} \beta_f \bar{D}_f  - \sum_\ell \mu_\ell\bar{C}^\pi_\ell + \sum_{\ell} \mu_\ell C^{av}_{\ell},
\end{align*}
where $\mu_\ell\geq 0$ is the multiplier associated with average link capacity constraint $\bar{C}^\pi_\ell \leq C^{av}_{\ell}$, and $\mu=\{\mu_\ell\}_{\ell\in E}$ is the vector containing these multipliers. We note that in the above, the policy $\pi$ under consideration is the primal variable. The dual function $D(\mu)$ is then given by,
\begin{align}\label{def:dual}
D(\mu)& = \max_{\pi}  \mathcal{L}(\pi,\mu) \notag\\
&= \left(\max_{\pi} \sum_{f} \beta_f \bar{D}_f  - \sum_\ell \mu_\ell\bar{C}^\pi_\ell\right)+\sum_{\ell}\mu_\ell C^{av}_{\ell},
\end{align}
where we note that only the term within the braces $\left(\cdot\right)$ depends on the policy $\pi$.

\subsection{Decentralized computation of Dual Function $D(\mu)$} \label{subsec:dualfun}
In order to evaluate the dual function at value $\mu$, the following problem needs to be solved,
$$
\max_{\pi} \sum_{f} \beta_f \bar{D}_f  - \sum_\ell \mu_\ell\bar{C}^\pi_\ell.
$$
Next, we make the following important observation. The cost $\sum_{\ell} \mu_\ell\bar{C}^\pi_\ell$ as well as the reward $\beta_f \bar{D}_f$ is the sum of the individual costs incurred by packets, i.e., 
\begin{align}
\sum_{\ell}\mu_\ell\bar{C}^\pi_\ell &=\sum_{\ell} \mu_\ell\sum_f \bar{C}^\pi_{\ell,f}\notag\\ 
&=\sum_f \left( \sum_{\ell} \mu_{\ell}\bar{C}^\pi_{\ell,f} \right)\notag\\
& = \sum_f \sum_{\sigma_f} \left( \sum_{\ell} \mu_{\ell}\bar{C}^\pi_{\ell,\sigma_f} \right),\label{eq:decom}
\end{align}
where $\bar{C}^\pi_{\ell,f}$ is the average bandwidth consumption by packets belonging to flow $f$ on link $\ell$, the index $\sigma_f$ labels packets of flow $f$, and $\bar{C}^\pi_{\ell,\sigma_f}$ denotes the average amount of link $\ell$ bandwidth consumed by packet $\sigma_f$. A similar decomposition holds for the timely-throughput reward too. This decomposition property yields us the following algorithm to compute the dual function $D(\mu)$.

\subsection{Highly Decentralized Packet Level Policy}\label{subsec:plp}
In this section, we will fix the value of dual variable at $\mu$, and focus exclusively on maximizing the following ``cumulative reward" earned by a policy $\pi$ in a decentralized way, 
\begin{align}\label{eq:cumrew}
 \sum_f \sum_{\sigma_f} \left( \beta_f \bar{D}_{\sigma_f} - \sum_{\ell} \mu_{\ell}\bar{C}^\pi_{\ell,\sigma_f} \right),
\end{align}
where $\bar{D}_{\sigma_f}$ denotes the probability that packet $\sigma_f$ is delivered to its destination node within its deadline. Maximization of the cumulative reward will yield us the value of dual function $D(\mu)$. 

\emph{Maximizing~\eqref{eq:cumrew} using Dynamic Programming } :\\
The \emph{state} $X(t)$ of the system at time $t$ is mentioned by describing the flow $f$ and age for each packet present at each node $i\in V$. We can then solve for the optimal $\pi$ that maximizes the cumulative reward~\eqref{eq:cumrew} using Dynamic Programming i.e., 
\begin{align}\label{eq:hjb}
R + V(x) = \max_{u} \left( R_{inst}(x,u) + \mathbb{E}_{u}\left\{V(y) \right\}    \right),
\end{align}
where $R$ is the optimal average reward, $V(x)$ is the transient reward function associated with the system beginning in state $x$, and $R_{inst}(x,u)$ is the one-step reward earned when the system state is $x$, and control $u$ is applied. The instantaneous reward $R_{inst}$ includes the reward earned due to timely delivery of packets, and the cost paid due to using the link bandwidth, i.e., $\mu_{\ell}$ amount of price is incurred upon using unit amount of link $\ell$ bandwidth. Solving the Dynamic programming equation~\eqref{eq:hjb}, and implementing the resulting policy leads to several technical difficulties:
\begin{itemize}
\item The number of variables involved in solving~\eqref{eq:hjb} is equal to the size of the state space. If we assume that the total number of packets in the network is bounded by $B$, the state space size is exponential in $B$ (one has to mention the location and age of each packet present in the network).  
\item The optimal policy calls for a centralized controller in order to be implemented, i.e., the control input at time $t$, $U(t)=\pi^\star(X(t))$ is a function of the system state $X(t)$. Thus, the nodes need to share their information with all the other nodes in every time-slot.
\end{itemize}

Our key result is that the cumulative reward can be maximized by maximizing the cumulative rewards earned by each individual packets.

We re-collect the decomposition principle~\eqref{eq:decom}, which says that the instantaneous reward asociated with the cumulative reward function~\eqref{eq:cumrew} is the sum of rewards of individual packets, i.e.,
\begin{align}\label{cost:pktlevel}
\sum_f \sum_{\sigma_f} \left(\sum_t \beta_f D_{\sigma_f}(t) + \sum_{\ell} \mu_\ell C_{\ell,\sigma_f}(t) \right),
\end{align}
where $D_{\sigma_f}(t)=1$ only if the packet $\sigma_f$ is delivered at time $t$ to its destination node $d_f$, and is $0$ otherwise, while $C_{\ell,\sigma_f}(t)$ is the amount of bandwidth utilized by the packet $\sigma_f$ at time $t$ on link $\ell$.

Since the total cost decomposes into the cost incurred by individual packets~\eqref{cost:pktlevel}, and since the reward of an individual packet can be optimized independently of other packets, it then follows that the cumulative reward~\eqref{eq:cumrew} can be optimized by implementing a ``packet-by-packet optimal policy". Thus, we introduce the following MDP which is concerned with optimizing the trajectory of a single packet from its source to destination.

\emph{Single Packet Optimal Transportation Problem} : 
Consider the following dynamic optimization problem. At time $t=0$, a single packet is generated at its source node $s_f$. Thereafter its evolution is jointly decided by the scheduling action applied at each of the node it traverses, and the prevailing channel state. Thus, if its transmission is attempted on a link $\ell$ at any time $t$, then the transmission succeeds with a probability $p_\ell$ which is the reliability of link $\ell$. Moreover, the link $\ell$ that is utilized for transmision charges a price of $\mu_{\ell}$ from the packet. After a sequence of transmissions occurring at different nodes $i\in V$, if the packet manages to reach the destination $d_f$ before time $\tau$, then it earns a reward of $\beta_f$ units.

The problem is to design a scheduling policy so as to maximize the net reward earned while transporting a unit packet from source to destination. In order to do so, we realize that the state of the packet at time $t$ is described by the node $i$ at which it is present, and its age, i.e. the time that has elapsed since it was generated at time $t=0$ at the source node. Solving the following DP equations yields the solution to the \emph{Single Packet Optimal Transportation Problem},
\begin{align}\label{eq:rl}
&V(i,s) = \max_{\ell=(i,j)}\left(\max_{u} \left\{ \mu_{\ell} u + P(\ell,u) V(j,s+1) \right.\right.\notag\\
&\left.\left.~~+ (1-P(\ell,u))V(i,s+1)       \right\}\right),
\end{align}   
where $u\in \{0,1\}$ represents the amount of bandwidh utilized for transmission, i.e., $u=0$ for not transmitting, and $u=1$ for transmitting. The joint action comprising the decisions $(\ell,u)$ ranges over all the choices of a transmission link $\ell$, or not transmitting the packet at all. We will denote the optimal policy thus obtained by solving the above DP as $\pi^{\star}_{f}(\mu)$, with the subscript $f$ denoting that the solution depends upon the flow $f$ that the individual packet $\sigma_f$ belongs to.

The size of the state space involved in solving the Single Packet Transportation Problem is equal to the number of nodes in the network $|V|$ times the deadline threshold $\tau$, i.e., $|V| \tau$. Moreover, the optimal decision for a packet $\sigma_f$ at any time $t$ depends only on its state, i.e., its age and location. Thus, it can be implemented in a distributed fashion, i.e., the node $i$ at which the packet is present simply looks up the optimal action to be taken, and implements it. It does not need to know the state of packets present at other nodes, or even the states of other packets present at the node $i$. Notice that this was not the case in implmenting the solution to~\eqref{eq:hjb}. 
\begin{lemma}\label{lemma:dualdec}
The policy $\pi^{\star}(\mu)$ that maximizes the Lagrangian $\mathcal{L}(\pi,\mu)$, or equivalently satisfies $\mathcal{L}(\pi,\mu) = D(\mu)$ implements the solution of the \emph{Single Packet Optimal Transportation Problem} for each packet $\sigma_f$ of each flow $f$. Thus, we have $\pi^{\star}(\mu)= \otimes_{f} \pi^{\star}_{f}(\mu)$.
\end{lemma}
\subsection{Obtaining the optimal prices  $\mu^\star$}\label{subsec:mu}
In the previous section, we derived an algorithm that computes the value of dual function $D(\mu)$, and $\pi^{\star}(\mu)$, i.e, the policy that maximizes the Lagrangian $\mathcal{L}(\cdot,\mu)$. However, in order to solve the dual MDP corresponding to the CMDP~\eqref{eq:intavg1}-\eqref{eqintavg13}, we need to solve the following dual problem,
\begin{align}\label{eq:dualprob}
\min_{\mu\geq 0} D(\mu).
\end{align}  
We will use sub-gradient descent method in order to converge to optimal link-prices $\mu^\star$. In the below, $k$ denotes the iteration index, and $\bar{C}^{\pi^{\star}(\mu)}_{\ell}$ denotes the average bandwidth consumption on link $\ell$ under the application of policy $\pi^{\star}(\mu)$ that can be calculated by solving the DP equations~\eqref{eq:rl} for each flow $f$.

In order to implement sub-gradient descent algorithm, each link $\ell$ needs to iterate on its price $\mu_{\ell}\ell(t)$ as follows
\begin{align}\label{eq:priceupd}
\mu_{\ell}(t+1) =\Pi\left[ \mu_{\ell}(t) + \alpha(t) \left(\bar{C}^{\pi^{\star}(\mu(t))}_{\ell}-C^{av}_{\ell}\right)  \right], t=1,2,\ldots,
\end{align}
\footnote{we have used the same index $t$ to label the time-slots, and the sub-gradient descent iterations. }where $\Pi\left[\cdot\right]$ projects the iterates onto a suitable compact set. Since the dual problem~\eqref{eq:dualprob} is convex, we have, 
\begin{figure}[!t]
	\centering
	\includegraphics[width=0.55\textwidth]{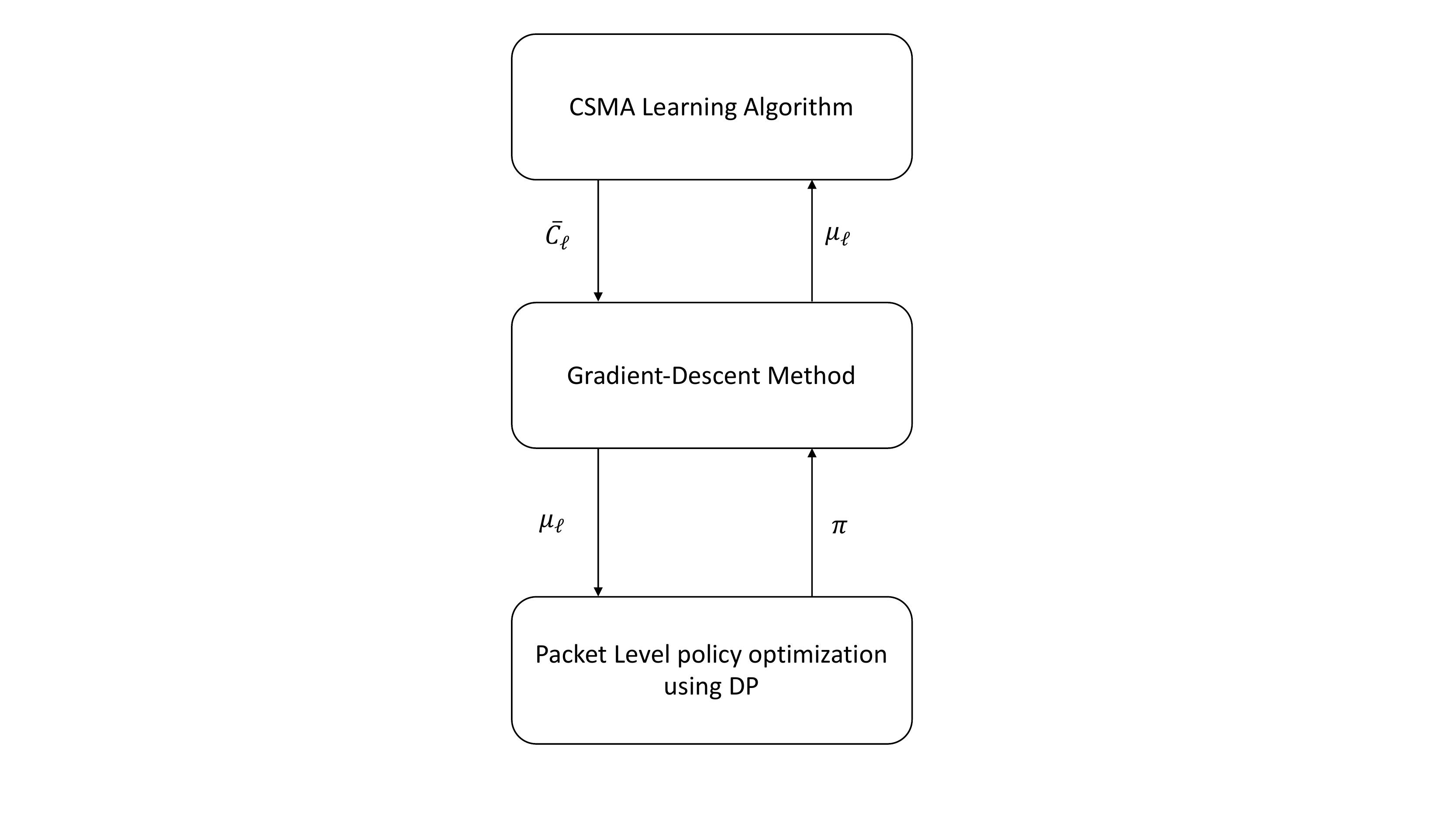}
	\caption{A two-layered iterative algorithm that solves the Policy Optimization Problem. Notice that the link-price tuner requires the value of average bandwidth congestion on each link $\ell$ that results under the application of policy $\pi(\mu)$.}
	\label{fig:1}
\end{figure}
\begin{lemma}
The price iterations~\eqref{eq:priceupd} converge to the price vector $\mu^\star$ that solves the dual problem~\eqref{eq:dualprob}.
\end{lemma}
We note that in order to carry out the price iterations we need to compute the quantities $\bar{C}^{\pi^{\star}(\mu(t))}_{\ell}$. This task is computationally expensive, and moreover, the assumption that there is a central entity that has knowledge of the network characteristics, is an unrealistic one. We provide a distributed scheme in the theorem below.
 
We summarize the results obtained in this section by concluding with the following Theorem.
\begin{theorem}[Scheduling under Average Link Bandwidth constraints]\label{th:linklevel}
Consider the problem of optimal scheduling for packets having end-to-end deadline constraints under link-level average constraints $\bar{C}_{\ell}^\pi \leq C^{av}_{\ell}, \forall \ell \in\mathcal{E}
$, i.e., the problem~\eqref{eq:p1}-\eqref{eq:c1}. The optimal policy is given by $\pi^{\star}=\otimes_{f}\pi_f^{\star}(\mu^{\star})$, where $\mu^\star$ is the solution to the dual problem~\eqref{eq:dualprob}. It implements the policy $\pi_f^{\star}(\mu^{\star})$ that is the solution to the single packet transportation problem with link prices set to $\mu^\star$, for each packet belonging to flow $f$. Hence in order to make decisions regarding a packet present at a node $i\in V$, the node $i$ only needs to know the age of the packet. 

The vector $\mu^\star$ of optimal prices can be obtained by performing the gradient descent iterations~\eqref{eq:priceupd}.In between two successive updates of the price $\mu(t)$, the DP iterations~\eqref{eq:rl} can be performed with price set to $\mu(t)$ in order to evaluate the quantity $\bar{C}^{\pi^{\star}(\mu(t))}_{\ell}$. This involves a link $\ell$ to obtain the value of value function evaluated at its outgoing links, i.e., if $\ell=(i,j)$ then all nodes $j:(j,k)\in\mathcal{E}$ need to share $V(j,\cdot)$. Hence, the price updates and value iterations can be performed in a distributed way.
\end{theorem}

 In practice, the iterations need to be performed using the data that is available during the network operation. Thus, in Section~\ref{sec:3layer} we briefly discuss a stochastic approximation based scheme which is an ``online learning" algorithm that guarantees convergence to the optimal policy.  
\begin{remark}
It must be noted that in this section we have addressed only a sub-problem concerning the relaxed version~\eqref{eq:intavg1}-\eqref{eqintavg13} of the timely throughput maximization problem. Thus, the following must be noted,
\begin{enumerate}
\item Since the policy $\pi^{\star}(\mu^{\star})$ is designed to satisfy link-level bandwidth constraints only on an average, its instantaneous bandwidth consumption might exceed $C^{av}_\ell$, i.e., $C^{\pi^{\star}(\mu^{\star})}_\ell(t)>C^{av}_\ell$ is a possibility.
\item As mentioned in Section~\ref{sec:rp}, we have not considered the problem of \emph{channel access} yet. Thus, so far we have explicitly assumed that interfering links have been provided orthogonal channels amounting to unlimited bandwidth via some mechanism. That is to say, at each time $t$, the set of available orthogonal sub-channels is allocated amongst the various network links $\ell\in E$ in such a manner that any two interfering links $\ell,\hat{\ell}\in \mathcal{E}$ are allocated sub-channels that are orthogonal. Furthermore there is no bandwidth constraint on these orthogonal channels. The problem of designing a decentralized channel access mechanism is highly non-trivial, and will be addressed in Section~\ref{sec:csma}.
 \item We have not addressed the constraint that the cumulative bandwidth consumed by the network at any time $t$ should be less than or equal to $K$ units. 
\end{enumerate}
In Appendix~\ref{sec:bo} we apply the gradient descent algorithm in order to further optimize over the vector of bandwidth allocation $\bar{I}$, and hence solve the problem~\eqref{eq:intavg1}-\eqref{eqintavg13}. The scheme discussed therein is impractical because it assumes that there is a centralized controller, and the scheme involves tuning the bandwidth allocated to each independent set in the set $\mathcal{I}$. The number of independent sets grows exponentially with the number of links $|\mathcal{E}|$, the scheme cannot be justifed for practical purrposes. This brings us to the CSMA protocol. 
\end{remark}

\section{CSMA for Decentralized Channel Access}\label{sec:csma}
We briefly discuss the CSMA protocol that we will utilize in order to obtain decentralized channel access. The CSMA scheme used by us is significntly different from the commonly used CSMA, and the differences will be pointed out at the end of this section. 

\subsection{Randomized Channel Access} We begin with a brief discussion of the CSMA protocol. We will now slightly augment the discrete time-slot model introduced earlier, in order to accomodate the CSMA scheme to be implemented in conjuntion with a scheduling policy. Thus, we will now assume that a small portion of each time-slot is devoted to making channel access decisions. We will call this dedicated time duration within each time-slot as a ``minislot". Thus, a time-slot is divided into a \emph{mini-slot} and a \emph{data-slot}, with the former reserved for making channel access decisions, and the latter for packet transmissions. At the beginning of each minislot, each link $\ell\in\mathcal{E}$ waits for a random amount of time duration that is exponentially distributed with mean value of $1\slash r_\ell$. The quantity $r_\ell$ is called the aggression parameter of link $\ell$. We will denote this random wait-time as counter. During a minislot, each link $\ell\in\mathcal{E}$ continually senses the carrier in order to detect packet transmissions from any of its conflicting links. 
At the expiry of its counter, if the link $\ell$ finds that none of its conflicting links\footnote{a link $\hat{\ell}$ such that $(\ell,\hat{\ell})$ is an edge in the edge interference graph.} is transmitting, then it makes the decision to attempt a packet transmission in the current data-slot.   

Since the support of an exponential random variable is the entire real line, we will truncate the wait counters to some large enough threshold value so that the probability that a counter value exceeds this threshold is vanishingly small. Thus, it is assumed that the duration of a minislot is much longer than the average value of waiting time, $1/r_{\ell}$. Under the above assumptions, the following fact is easily verified.
\begin{lemma}\label{eq:csmaband}
Under the above described randomized channel access scheme, the probability that a link $\ell\in\mathcal{E}$ gets channel access to transmit a packet in a data-slot $t$ is given by,
\begin{align}\label{eq:adhoc4}
p(\ell;r) = \frac{r_\ell}{\sum_{\hat{\ell}\in N(\ell)}r_{\hat{\ell}}},
\end{align}
where the vector $r:=\left(r_1,r_2,\ldots,r_{|\mathcal{E}|}\right)$, and $N(\ell)$ is the set of links $\hat{\ell}$ that interfere with the link $\ell$.  Equivalently, the bandwidth available to link $\ell$ under the CSMA-$r$ protocol is equal to $\frac{r_\ell}{\sum_{\hat{\ell}\in N(\ell)}r_{\hat{\ell}}}$.
\end{lemma}
We will denote the above randomized channel access mechanism as CSMA-$r$. The CSMA algorithm is decentralized because each link carries out sensing and channel access independently of other links in the network, and hence it does not require a centralized co-ordinator to ensure that average bandwidth constraints are satisfied. We will use $p(r):=\{p(\ell;r)\}_{\ell\in\mathcal{E}}$ to denote the vector consisting of average link bandwidths under the CSMA-$r$ protocol.
\begin{remark}
We note that the CSMA model considered by us is significantly different from the existing commonly used CSMA scheme as in~\cite{liew2010back,wang2005throughput}. Under the commonly used CSMA, the set of links active at any time $t$, is described by a Markov process, with its state-space equal to $\mathcal{I}$, i.e., the set of independent sets of the link interference graph.  However in our set-up, the set of active links, i.e., $IS(t)$, is i.i.d. across each time-slot $t=1,2,\ldots$. Such a construction is required because we need to guarantee that the temporal bandwidth fluctuations are minimal, which is necessary in order to ensure optimality of scheduling policy with respect to timely-throughput metric. More concretely, we cannot allow for large amount of fluctuations in link-bandwidths. Thus, for example, the contribution of a single packet to the timely throughput is an intricate function of the bandwidth availability across various links $\ell\in\mathcal{E}$ \emph{over a time horizon of $\tau$ time-slots}, which is the time that the packet spends in the multi-hop network. This is in contrast with the network queue stability problems, where temporal fluctuations in bandwidth availability do not affect the throughput as long as the average link-bandwidth remains the same~\cite{tassi1}. The primary reason why such a control on bandwidth fluctuation is required, is because the ``current utility" of a packet depends on its ``age", i.e., the time it has spent in the network.   
\end{remark}
\section{Capacity Scaling and Asymptotically Optimal Policy}
We will develop a decentralized scheduling policy and show that is asymptotically optimal if the cumulative networkwide-available bandwidth is scaled to $\infty$. In order to develop the policy, we will combine the solution of the relaxed problem with the CSMA protocoland then analyze its timely throughput in this limiting regime. We begin by formally defining the network scaling that we employ. 

Recall the definition of $f(C^{av})$ as in~\eqref{eq:p1}
\begin{align}
f(C^{av}) := \max_{\pi: \bar{C}^\pi\leq C^{av}} \sum_f \beta_f \bar{D}_f.
\end{align}
Now, assume that the packet arrivals for each flow $f$ are random and follow the Bernoulli distribution with parameters $\left(1,A_f\right)$. 
We now introduce a performance metric that is similar to $f(C^{av})$. Let us assume that at the beginning of each time-slot $t=1,2,\ldots$, each link $\ell\in \mathcal{E}$ is now available with a probability $p(\ell;r)$, which is the activation probability of link $\ell$ under the CSMA-$r$ protocol~\eqref{eq:adhoc4}. The stochastic availability of the links is used to model the random activations of links by the CSMA scheme. Now define
\begin{align}\label{eq:ftilde}
\tilde{f}_{1}(p(r)) := \max_{\pi: C^\pi(t)\leq C^{CSMA_r}_{1}(t)} \sum_f \beta_f \bar{D}_f,
\end{align}
where $C^{CSMA_r}_{1}(t)=\{C^{CSMA_r}_{1,\ell}(t)\}_{\ell\in\mathcal{E}}$ is the vector of bandwidths allocated at time $t$ under the CSMA-$r$ protocol applied to a unit bandwidth. The subscript $1$ in the above stands for the fact that only a single independent set is to be activated at each time-slot $t$.

Let us now consider a sequence of wireless networks. For the $N$-th network in the sequence, we have that the packet arrivals for each flow $f$ are distributed according to Bernoulli $\left(N,A_f\right)$. Also, the network has access to $N$ units of bandwidth, and hence can now activate $N$ independent sets simultaneously in any time-slot $t$. The channel access mechanism in the $N$-th network is as follows. The network uses $N$ \emph{independent} CSMA counters for channel access on $N$ orthogonal channels. Each link $\ell$, at the beginning of each mini-slot, generates $N$ i.i.d. backoff counters which are exponential with mean $1\slash r_\ell$, one for each orthogonal channel. Then, it uses a single counter on the corresponding channel in order to apply the CSMA scheme on it. Hence, the number of orthogonal channels available to link $\ell$ at each time $t$ is distributed according to Bernoulli $(N,p(\ell;r))$. Define 
\begin{align}\label{eq:ftildeN}
\tilde{f}_{N}(p(r)) := \max_{\pi: C^\pi(t)\leq C_{N}^{CSMA_r}(t)} \sum_f \beta_f \bar{D}_f,
\end{align}
where the sub-script $N$ in $C_N^{CSMA_r}(t):=\{C_{N,\ell}^{CSMA_r}(t)\}$ denotes that the network has $N$ orthogonal channels available to it, and superscript $r$ denotes that independent CSMA counters with aggression parameter $r$ are used on each of them separately.

We note that the quantity $\tilde{f}_N(p(r))$ is less than or equal to $f(Np(r))$ because of the following observation.  The set of policies that qualify while evaluating $f(Np(r))$ have no constraint on instantaneous bandwidths during individual time-slots $t=1,2,\ldots$. However during the computation of $\tilde{f}_{N}(p(r))$, the set of allowable policies can utilize only $C^{CSMA_{r}}_{N}(t)$ amount of links at time $t$. While since under the CSMA scheme, the infinite horizon average bandwidth consumption is equal to $Np(r)$, the set of policies that are feasible during the evaluation of $\tilde{f}_N(p(r))$ are automatically feasible for the evaluation of $f(Np(r))$.  
\begin{lemma}\label{lemma:capacityineq}
We have,
\begin{align}
\tilde{f}_N(p(r)) \leq f(Np(r)).
\end{align}
\end{lemma}
\begin{remark}
It is this hard, per time  restriction on the available bandwidth that makes the optimal scheduling problem for CSMA network much more challenging, since a scheduler now has to prioritize amongst the packets based on the global state of the network, thus requiring a centralized controller. However, as will be shown now, it is possible to overcome this limitation if the optimal prices $\mu^\star$ are utilized appropriately while making scheduling decisions.
\end{remark}
Next, we show that the relative difference between $f(Np(r))$ and $\tilde{f}_N(p(r))$ asymptotically vanishes as the network capacity is scaled to $\infty$, i.e., $\frac{f(Np(r)) - \tilde{f}_{N}(p(r))}{f(Np(r)) }\to 0$ as $N\to\infty$. Hence, asymptotically nothing is lost due to restraining the link capacities to those made available by the CSMA algorithm. Our proof relies on constructing a decentralized scheduling algorithm for the CSMA network, denoted $\tilde{\pi}$, for which the timely-throughput is within $O\left(\sqrt{N}\right)$ of $f(Np(r))$. We now describe our scheduling algorithm $\tilde{\pi}$.

Let $\pi^{\star}$ denote the policy that is optimal for the scheduling problem under the link-level average bandwidth constraints given by $p(r)$, i.e., $\pi^\star$ solves the problem~\eqref{eq:p1}-\eqref{eq:c1} with $C^{av}$ set equal to $p(r)$. $\pi^\star$ can be obtained as in Theorem~\ref{th:linklevel}.

\emph{Construction of $\tilde{\pi}$}: 
It follows from Theorem~\ref{th:linklevel} that the policy $\pi^{\star}$ makes packet-based decisions at each node $i\in V$. Since $\pi^{\star}$ does not make decisions based on instantaneous bandwidth availability $C^{CSMA_r}_{N}(t)$, it is not a feasible policy for scheduling under the CSMA protocol applied to $N$ orthogonal channels. Thus, one cannot utilize $\pi^{\star}$ in order to schedule packets for the $N$-th scaled network. 

Now, if at some timeslot $t$ it occurs that according to $\pi^{\star}$ a node $i$ has to utilize more than $C^{CSMA_r}_{N,\ell}(t)$ 
amount of bandwidth on a link $\ell$, then the node $i$ simply chooses a maximal subset of the packets meant for transmission on link $\ell$ subject to total bandwidth utilization less than $C^{CSMA_r}_{N,\ell}(t)$. The selection of the set of packets meant for transmission on link $\ell$ can be made according to some rule that has been fixed apriori before the network operation begins at time $t=0$. The policy $\tilde{\pi}$ is essentially $\pi^\star$ truncated according to $C^{CSMA_r}_{N}(t)$.
\begin{theorem}\label{th:scale}
Consider the sequence of ``scaled CSMA-$r$ networks" as defined above operating under the policy $\tilde{\pi}$. We then have that 
\begin{align}
\frac{f(Np(r)) - \tilde{f}_{N}(p(r))}{f(Np(r)) } = O\left(\frac{1}{\sqrt{N}}\right),
\end{align}
where $\tilde{f}_{N}(p(r))$ is the maximum timely-throughput attainable by the $N$-th CSMA network in the sequence.
\end{theorem}
\begin{IEEEproof}
In the below, we drop the reference to the scale $N$, and the CSMA aggression vector $r$, e.g. $C_N^{CSMA_r}(t)$ becomes $C^{CSMA}(t)$.

The following arguments are based on analysis of the evolutions of policies on an appropriately constructed probability space. Let us denote by $r_0$ the (average) reward earned by policy $\pi^{\star}$ under the average bandwidth constraint on link $\ell$ equal to $NC(r)$. Firstly note that the reward collected by the policy $\tilde{\pi}$ (denoted by $r_1$) does not increase if it were to, instead of dropping a packet because of violation of instantaneous capacity $C^{CSMA}_\ell(t)$, schedule it as dictated by $\pi^{\star}$, but no reward is given to it if this packet is delivered to its destination node (denoted by $r_2$).
 However $r_2$ is more than the reward if now a penalty of $\beta_f$ units per packet was imposed for scheduling a packet via utilizing ``capacity in excess of $C^{CSMA}(t)$" at some link $\ell \in \mathcal{E}$, but it were given a reward in case this packet reaches the destination node (denoted by $r_3$). 

$r_3$ is certainly more than the reward which $\pi^{\star}$ earns if it is penalized an amount equal to the sum of the excess bandwidths (in excess of $C^{CSMA}(t)$) that its links utilize  (denoted by $r_4$) multiplied by $\beta_f$, since any individual packet may be scheduled multiple times by utilizing excess bandwidth. Thus, the difference $r_0-r_4$ is less than the sum of the excess bandwidths utilized by the links operating under the policy $\pi^{\star}$, scaled by $\max_f \beta_f$'s. Next, we will derive a bound on the excess capacity utilization.

Consider the system operation under the policy $\pi^\star$.  Based on the above arguments, we thus have that, (let all $\beta_f\equiv 1$),
\begin{align}\label{ineq:1}
r_0-r_4 \leq \lim_{T\to\infty} \frac{1}{T}\mathbb{E}\sum_{t=1}^{T}\sum_{\ell\in\mathcal{E}} \left(C^{\pi^\star}_\ell(t)- C^{CSMA}_\ell(t)\right)^+ 
\end{align}
Note that 
\begin{align}\label{ineq:2}
&\left(C^{\pi^\star}_\ell(t)- C^{CSMA}_\ell(t)\right)^+\notag\\
&= \left( \left(C^{\pi^\star}_\ell(t)- C_\ell(r)\right) + \left(C_\ell(r) -C^{CSMA}_\ell(t)\right)\right)^+ \notag\\
&\leq  \left(C^{\pi^\star}_\ell(t)- C_\ell(r)\right)^{+} + \left(C_\ell(r) -C^{CSMA}_\ell(t)\right)^{+}.
\end{align}
We also note that 
\begin{align}\label{eq:1}
C^{\pi^\star}_\ell(t) = \sum_{f}\sum_{\tau}C^{\pi^\star}_{f,\ell,\tau}(t),
\end{align} 
where  $C^{\pi^\star}_{f,\ell,\tau}(t)$ denotes the bandwidth utilization at time $t$ on link $\ell$ by packets of flow $f$ that have an age of $\tau$ time-slots.
Similarly,
\begin{align}\label{eq:2}
C_\ell(r) = \sum_{f}\sum_{\tau}C_{f,\ell,\tau},
\end{align} 
where $C_{f,\ell,\tau}$ denotes the average bandwidth utilization on link $\ell$ by packets of flow $f$ that have an age of $\tau$ time-slots. Using~\eqref{eq:2},~\eqref{ineq:1} and~\eqref{ineq:2} we upperbound the term $\left(C^{\pi^\star}_\ell(t)- C^{CSMA}_\ell(t)\right)^+$ in the r.h.s. of~\eqref{eq:1} as,
\begin{align}
\left(C^{\pi^\star}_\ell(t)- C^{CSMA}_\ell(t)\right)^+  &\leq \left(\sum_{f,\tau}\left(C_{f,\ell,\tau}(t)- C_{f,\ell,\tau}\right)^+\right) \notag\\
&+ \left(C^{CSMA}_\ell(t)- C_{\ell}(r)\right)^+
\end{align}
Combining~\eqref{ineq:1} with the above, we obtain that  
\begin{align*}
& r_0-r_4 \\
&\leq \lim_{t\to\infty}  \frac{1}{T}\sum_{t=1}^{T}\sum_{f,\tau} MAD\left(C_{f,\ell,\tau}(t)\right) + MAD\left(C^{CSMA}_\ell(t)\right)\\
&= O(\sqrt{N}) + O(\sqrt{N})
\end{align*}
where $N$ is the scaling parameter for packet arrivals. Thus we have that,
\begin{align*}
r_o-r_4 \leq O\left(\sqrt{N}\right).
\end{align*} 
Since the quantity $f(Np(r))$ scales linearly with $N$, this completes the proof. 
\end{IEEEproof}
Next, we show that if the parameter $r$ of the CSMA schem is chosen appropriately so as to optimize the bandwidths allocated to various independent sets in $\mathcal{I}$  ``optimally", then the policy $\tilde{\pi}$ is also asymptotically optimal for the original problem. 

\begin{theorem}\label{eq:asymptot}
Let $OPT$ denote the value of the relaxed problem~\eqref{eq:intavg1}-\eqref{eqintavg13} that was obtained by relaxing the original timely-throughput maximization problem~\eqref{op:int1}-\eqref{op:int2}. There exists a value of the aggression parameter $r^\star$ such that for the CSMA $r^{\star}$ network operating under the policy $\tilde{\pi}$, we have that the timely throughput $\left(\sum_f \bar{D}_f\right)_{\tilde{\pi}}$ is greater than or equal to $OPT-O(\sqrt{N})$, i.e.,
$$
\frac{OPT - \left(\sum_f \bar{D}_f\right)_{\tilde{\pi}}}{OPT} = O\left(\frac{1}{\sqrt{N}}\right),
$$
and hence asymptotically CSMA $r^{\star}$ utilized in combination with $\tilde{\pi}$ is asymptoticaly optimal for the timely throughput maximization problem~\eqref{op:int1}-\eqref{op:int2}.
\end{theorem}
\begin{IEEEproof}
Let us denote by the set $\mathcal{S}$, the set of vectors that describe the \emph{instantaneous} average bandwidths available to each independent set. Thus,
\begin{align*}
\mathcal{S} &= \left\{ \bar{I}_m(t): \bar{I}_m(t)\mbox{ is average bandwidth available to}\right.\\
&\qquad \left. IS_m \mbox{ at time } t             \right\}.
\end{align*}
The closure of the set $\mathcal{S}$, i.e. $\bar{\mathcal{S}}$ then coincides with the set $\{\bar{I}:\bar{I}_m\geq 0,\sum_m\bar{I}_m = K   \}$. In particular, $\bar{I}^\star$, the time-average bandwidth that is optimal for the relaxed problem~\eqref{eq:intavg1}-\eqref{eqintavg13}, also lies in the set $\mathcal{S}$. Let $tp_1$ be the timely-throughput of the policy that attains the maximum while evaluating $f\left( C_\ell ( \sum_{m:\ell\in IS_m} \bar{I}^\star_m )\right)$. Also, let $tp_2$ the timely throughput of the scheduling policy that is optimal when applied in conjunction with the CSMA-$r^{\star}$, where the parameter $r^{\star}$ is chosen so that the expected bandwidths allocated at \emph{each time-slot} are given by $\bar{I}^\star$. Such an $r^{\star}$ exists because any allocation in the set $S$ can be obtained through an appropriate choice of $r$. It then follows from Theorem~\ref{th:scale} that the difference between $tp_1$ and $tp_2$ is $O(\sqrt{N})$, and hence asymptotically, as $N\to\infty$, the optimal throughput achievable under the CSMA protocol is the same as the solution of the relaxed problem
~\eqref{eq:intavg1}-\eqref{eqintavg13}.
\end{IEEEproof}
\begin{remark}
Utilizing multiple independent copies of CSMA  protocol allows us to smoothen the bandwidth fluctuations for a time duration equal to the deadline $\tau$, which is the time taken by a packet to reach its deadline. This helps us in ensuring that a single packet that is generated during the time-slot $s$, views the link-capacities $\{C_{\ell}(t)\}_{\ell\in\mathcal{E}},t\in [s,s+\tau]$ as nearly equal to their average values $\{\bar{C}_{\ell}\}$ \emph{for its entire lifetime in the network}. However, since the number of packets generated by the network also scales linearly in $N$, hence it is not trivial, in the light of utilizing a complicated policy such as $\tilde{\pi}$, to ensure that a packet receives its ``right share of bandwidths" over its entire lifetime, one that ensures that the timely throughput is not affected.
\end{remark}
\section{Obtaining $r^{\star}$} \label{subsec1}
Though Theorem~\ref{eq:asymptot} ensures the existence of an $r^{\star}$ such that the combination of CSMA $r^{\star}$ and $\tilde{\pi}$ can be used to attain the network timely throughput capacity in a decentralized fashion, it does not discuss how to obtain $r^{\star}$. Since obtaining the performance of $\tilde{\pi}$ as a function of CSMA parameter $r$, and the scaling parameter $N$ is a difficult problem, we will instead optimize the timely throughput under the average bandwidth constraint~\eqref{eq:p1}-\eqref{eq:c1}.

Consider the following problem, dubbed the~\emph{CSMA Optimization Problem}.
Define
\begin{align}
F(r) := \sup  \left\{f(C^{av}) : \mbox{ s.t. } C^{av}_\ell = \frac{r_\ell}{\sum_{\hat{\ell}\in N(\ell) } r_{\hat{\ell}}},\forall \ell \in E\right\},\label{eq:adhoc13}
\end{align}
CSMA optimization problem is
\begin{align}\label{eq:csmaopt}
\max_{r\geq 0} F(r).
\end{align}
We now turn our attention towards obtaining its solution $r^\star$ in a distributed manner.  
 
 Next, we compute the gradient of the function $F(\cdot)$ with respect to the CSMA aggression parameter $r$ which can be used in the gradient-descent method for optimizing the function $F(\cdot)$. In the below, we let 
\begin{align}\label{eq:cumaggr}
R_\ell = r_\ell + \sum_{\hat{\ell}\in N(\ell)}r_{\hat{\ell}},
\end{align} 
be the ``cumulative aggression" associated with link $\ell$ and its neighbouring links. 
The following results are easily derived.
\begin{lemma}\label{lemma3}
\begin{align*}
p(\ell;r) = \frac{r_\ell}{R_\ell},
\end{align*}
so that for a link $\hat{\ell}$ that interferes with link $\ell$, i.e., $\hat{\ell}\in N(\ell)$, we have that,
\begin{align}
\frac{\partial p(\ell;r)}{\partial r_{\hat{\ell}}}  = -\frac{r_\ell}{R_\ell^2},
\end{align}
while,
\begin{align}
\frac{\partial p(\ell;r)}{\partial r_{\ell}}  = \frac{R_\ell - r_\ell }{R_\ell^2}.
\end{align}
Hence it follows from Lemma~\ref{lemma:shadow} (see Appendix) that,
\begin{align}
\frac{\partial F}{\partial r_{\ell}} = \sum_{\hat{\ell}\in N(\ell)} \frac{-\mu^{\star}_{\hat{\ell}}(C(r))       }{R^2_{\hat{\ell}}} + \frac{\left(R_{\ell}-r_{\ell}\right)\mu^{\star}_\ell(C(r))}{R^2_\ell}
\end{align}
\end{lemma}
After having derived the explicit expressions for the gradients, we are now in a position to apply the gradient-descent scheme in order to optimize the CSMA-$r$ protocol,
\begin{theorem}\label{th:csma}
Consider the CSMA optimization problem~\eqref{eq:csmaopt}, with the function $F(\cdot)$ defined as in~\eqref{eq:adhoc13}.

Denote by $\mu^\star(C(r)):=\{\mu^{\star}_{\ell}(C(r))\}_{\ell\in\mathcal{E}}$ the value of link prices that solve the dual problem~\eqref{eq:dualprob} with average link bandwidth constraints set equal to $\frac{r_\ell}{\sum_{\hat{\ell}\in N(\ell)}r_{\hat{\ell}}}$ (see~\eqref{eq:adhoc4}). 
In the below, variable $k$ denotes the iteration index associated with ``$r$ updates". Consider the following iterative algorithm in which each link $\ell\in\mathcal{E}$ tunes its parameter $r_\ell$ according to (in the below, $\mu^k_\ell$ is to be read as $\mu^\star_\ell(C(r^k))$, similarly for $R^k_\ell$ etc.),
\begin{align}\label{eq:csmalearn1}
&r^{k+1}_\ell = \Gamma\left(r^k_\ell + \gamma^k  \left(\sum_{\hat{\ell}\in N(\ell)} \frac{-\mu^k_{\hat{\ell}} }{(R^k_{\hat{\ell}})^2} + \frac{\left(R^k_{\ell}-r^k_{\ell}\right)\mu^k_\ell}{(R^k_\ell)^2}\right)        \right), \notag\\
&\forall \ell \in E, k = 1,2,\ldots..
\end{align}
The above iterations converge to a locally optimal value of $r$ for the problem~\eqref{eq:csmaopt}. The complexity of this algorithm is $O(|\mathcal{E}|)$. 
\end{theorem}
\begin{remark}
Let us break down the various components involved in performing the iterations~\eqref{eq:csmalearn1}. An update of the $r^k_{\ell}$ involves access to the values of $\mu^\star_\ell(C(r^k))$ and the quantity $R^k_{\ell}$. Since $R^k_{\ell}=\sum_{\hat{\ell}\in N(\ell)}r^k_{\hat{\ell}}$, the quantity $R^k_{\ell}$ is easily available if we allow the links to share their values $r^k_{\ell}$ with their neighbours.

Now, the quantity $\mu^\star_\ell(C(r^k))$ can be computed by performing gradient-descent and value iterations as in Theorem~\eqref{th:linklevel} by setting the link bandwidths at $p(r)$ (or $C(r)$). This involves the nodes to share the values of value function with their neighbours. 

The overall scheme thus requires information sharing amongst neighbouring nodes only.
 \end{remark}

Section~\ref{sec:3layer} discusses the problem of searching the optimal policy using online learning methods that use data available during the operation of network.
\section{Simulation Results}\label{sec:simu}
We now carry out simulations to test the performance of the policy that was shown to be asymptotically optimal in Theorem~\ref{eq:asymptot}. However note that the CSMA modulator of Theorem~\ref{th:csma} that solves the CSMA optimization problem converges to an $r^\star$ that is only locally optimal. However, simulation results show that the resulting policy is quite good in practice.

We will refer to the policy simply as the ``optimal policy", with the understanding that it is using an $r^\star$ that may not be optimal, and also that even if the $r^\star$ were to be globally optimal, the policy of Theorem~\ref{eq:asymptot} is asymptotically optimal in the limit the network scale $N\to\infty$.

\subsection{Policy Description}\label{subsec:policy}
We compare the performance of the optimal policy with a version of the Q-CSMA policy~\cite{srikant1,wlrnd1,jiang2010distributed} that has been adapted to be relevant to the problem of maximizing the timely throughput. We denote this policy as \emph{Q-CSMA with EDF-Shortest Path}, which is described below.

\emph{Q-CSMA with EDF-Shortest Path}: The Q-CSMA algorithm~\cite{srikant1} has been shown to throughput optimal for wireless networks in which interference is modeled using edge intereference graph. The Q-CSMA algorithm uses a decentralized channel access mechanism in which the CSMA aggression parameter $r$ is tuned in accordance with the current queue lengths $Q(t)$. We describe a discretized version of the Q-CSMA algorithm that was discussed in~\cite{srikant1}. A single time-slot is divided into a control mini-slot and a data slot. It is during the control mini-slot, that decisions regarding channel access are made.  A single control mini-slot is divided into $W$ sub-slots. During the control mini-slot for time-slot $t$, each link $\ell$ generates a number $w_\ell(t)$ uniformly at random from the set $\{1,2,\ldots,W\}$. The quantity $W$ is called the window-size. The link $\ell$ then declares an ``intent" during the sub-slot $w_\ell(t)$ in case none of its neighbouring links have declared an intent by the sub-slot $w_{\ell}(t)$. At the end of the control mini-slot, if the link $\ell$ does not hear intent from any of its neighbours, and if none of its neighboring links were transmitting during data slot $t$, then the link $\ell$ transmits during the data-slot for time $t$ with a probability equal to $e^{Q_{\ell}(t)}/1+Q_{\ell}(t)$, where $Q_{\ell}(t)=\sum_f Q_{f,\ell}(t)$ is the cumulative queue length at link $\ell$. Also, if multiple neighboring links declare intent in the same sub-slot, then none of them transmits data during the corresponding data-slot.

Note that the Q-CSMA provides only channel access decisions, but not the routing or packet scheduling decisions that prioritize based on the age of packets. Thus, we will combine the Q-CSMA with EDF discipline which will enable it to make routing decisions, and also the earliest deadline first (EDF) policy which will allow it to prioritize the scheduling of packets that are ``closer" to their deadline. For such a policy, the bandwidth allocated across the network links during a data-slot $t$ are decided by the Q-CSMA algorithm described above. At the end of the control-slot, each link $\ell$ arranges the packets with it in increasing order of their time until deadline. Then, the link $\ell$ schedules them on the shortest path route, subject to the instantaneous link bandwidth of link $\ell$ that has been provided to it by the Q-CSMA. In case there are multiple shortest paths that connect the link $\ell$ to the destination node for a flow $f$, then the link $\ell$ chooses from amongst them uniformly at random while scheduling packets for flow $f$.
\begin{figure}[h]
	\centering
	\includegraphics[width=0.4\textwidth]{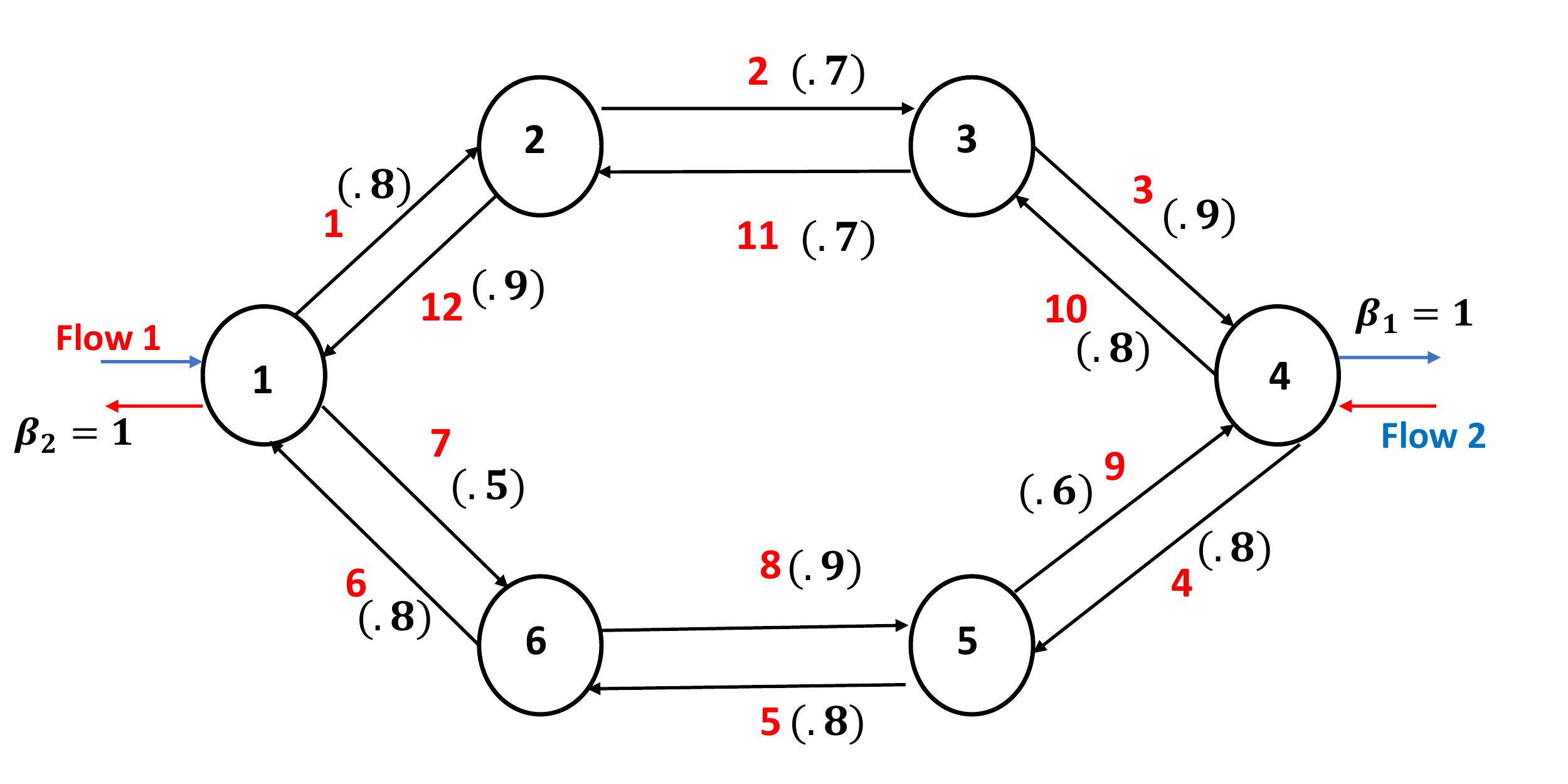}
	\caption{A multihop wireless network shared by two flows. The source-destination pairs are $(1,4)$ and $(4,1)$. Channel reliabilities for the links are provided within braces, eg. reliability of link $(2,3)$ is equal to $.7$. It is assumed that any links that share a node interfere with each other. }
	\label{fig:simutopo1}
\end{figure}
\vspace{-.5cm}
\begin{figure}[h]
	\centering
	\includegraphics[width=0.4\textwidth]{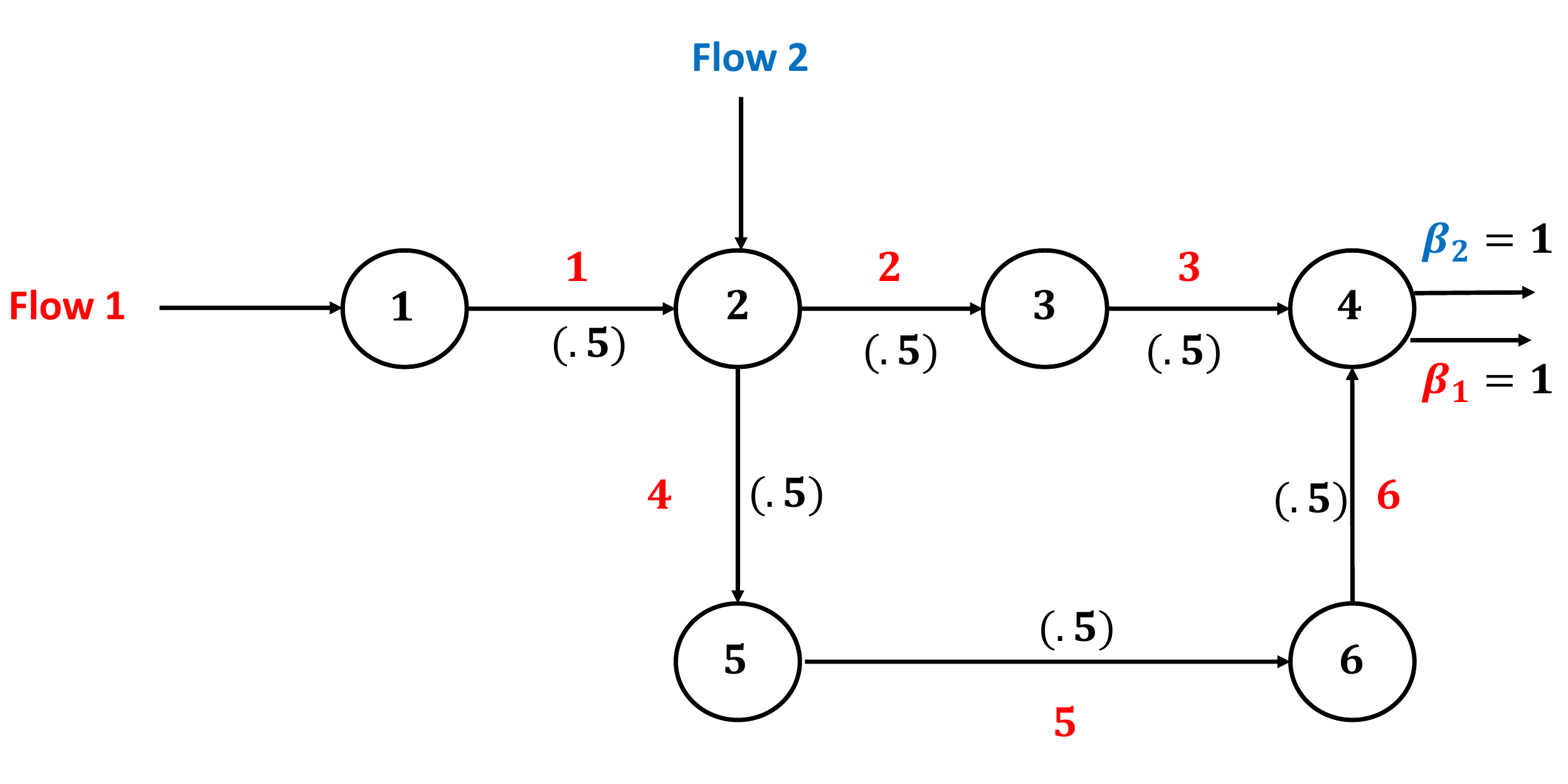}
	\caption{A multihop wireless network shared by two flows with the source-destination given by $(1,4)$ and $(2,4)$. Channel reliabilities for the links are provided within braces, and is equal to $.5$ for all of the network links. It is assumed that any two links that share a node interfere with each other. The corresponding link interference graph is shown in Fig.~\ref{fig:linkinterferencegraph}.}
	\label{fig:simutopo2}
\end{figure}
\begin{figure}[h]
	\centering
	\includegraphics[width=0.22\textwidth]{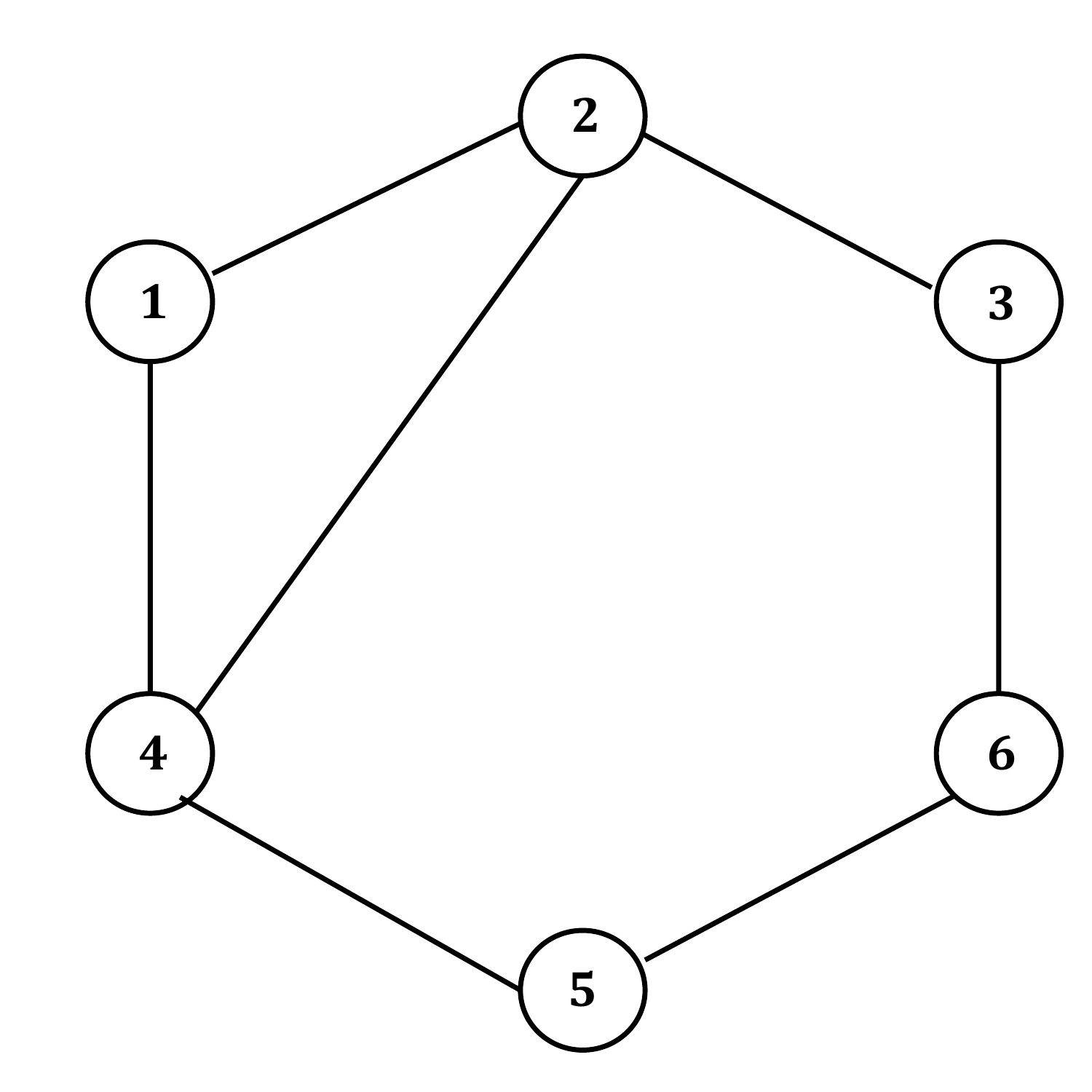}
	\caption{Link interference graph for the network of Fig.~\ref{fig:simutopo2}.}
	\label{fig:linkinterferencegraph}
\end{figure}
\subsection{Network Set-Up}\label{subsec:setup}
We simulate the policies for the networks shown in Fig.~\ref{fig:simutopo1} and Fig.~\ref{fig:simutopo2}. We assume that any two links which share a node will be affected by wireless interference, and hence will be connected by an edge in the link interference graph. Thus, two links $\ell_1 = (i_1,j_1)$ and $\ell_2=(i_2,j_2)$ interefere if either of the following conditions is satisfied $i_1=i_2,i_1=j_2,j_1=i_1,j_1=j_2$. Throughout, we assume that all links have a transmission capacity of $1$ pkt/time-slot. For the Q-CSMA with EDF-SP policy, we set the window length of the control mini-slot to be equal to $10$ sub-slots. We assume that for the unscaled network, the arrivals for each flow $f$ at each time $t$ are distributed according to Bernoulli $(1,.8)$.
\subsection{Results}\label{subsec:results}
\begin{figure}[h]
	\centering
	\includegraphics[width=0.4\textwidth]{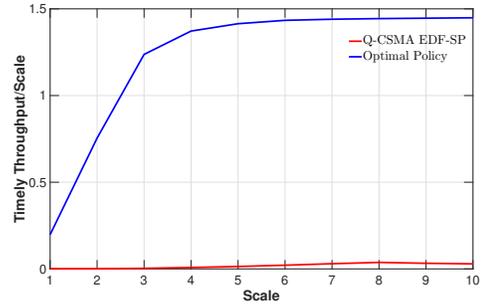}
	\caption{A plot of the normalized timely throughput for the network of Fig.~\ref{fig:simutopo1} as the scale of the network is varied. We notice that the scaled throughput equilibriates quite ``fast" at a scaling of $4$, hence hinting that the sub-optimality bounds of $O(\frac{1}{\sqrt{N}})$ derived in Theorem~\ref{eq:asymptot} might be pessimistic, and might be further improved upon.}
	\label{fig:tps1}
\end{figure}

\begin{figure}[h]
	\centering
	\includegraphics[width=0.4\textwidth]{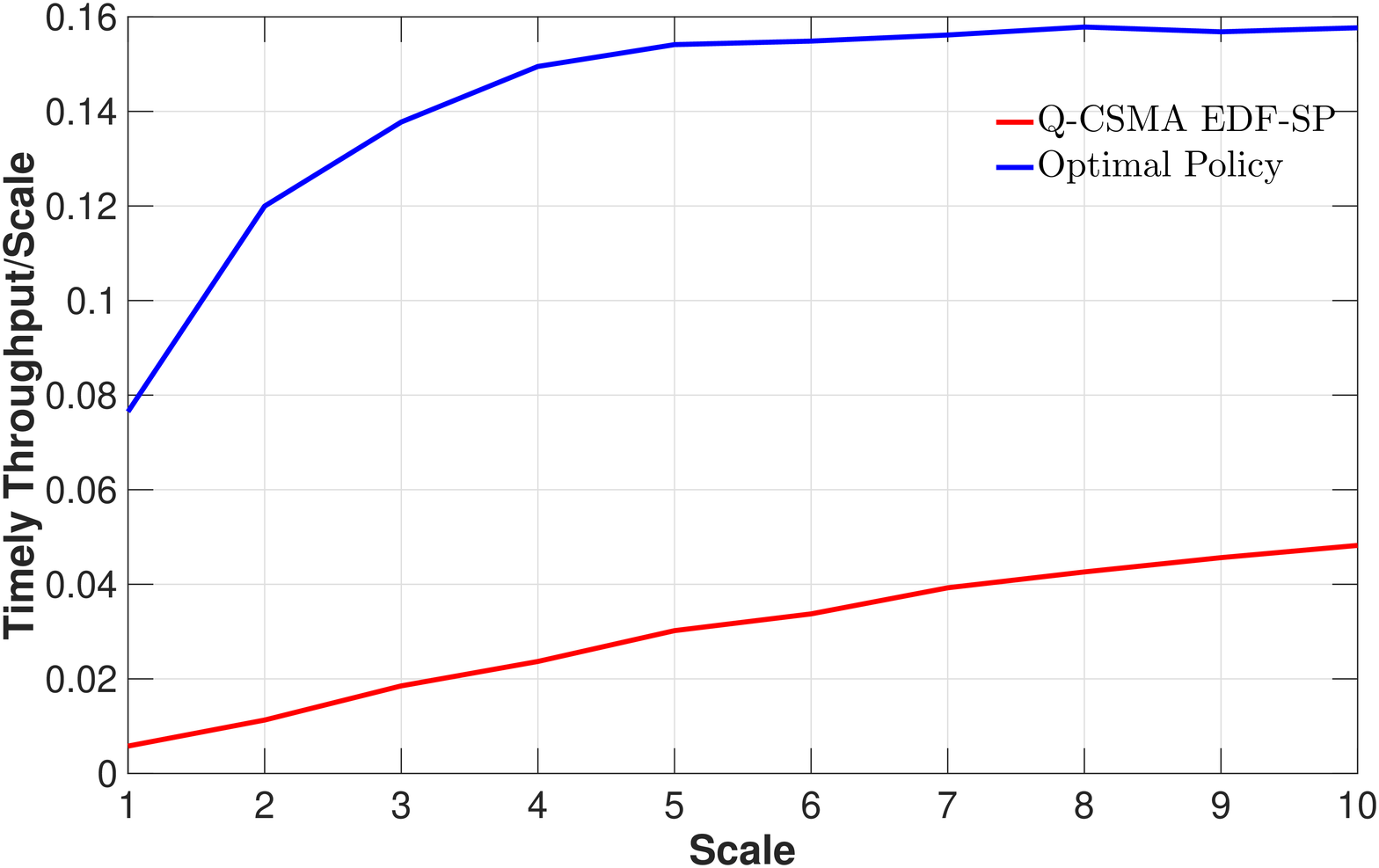}
	\caption{A plot of the normalized timely throughput for the network of Fig.~\ref{fig:simutopo2} as the scale of the network is varied. Similar to the observation made in the plot of Fig.~\ref{fig:tps1}, we notice that the scaled throughput equilibriates quite ``fast" at a scaling of $4$. Thus, the sub-optimality bounds of $O(\frac{1}{\sqrt{N}})$ derived in Theorem~\ref{eq:asymptot} might be possibly further improved upon.}
	\label{fig:tps2}
\end{figure}

We fix the relative end-to-end deadline for the flows to be equal to $10$ time-slots, and vary the network scale $N$ of Theorem~\ref{th:scale}. 
The resulting timely throughputs are plotted in Fig.~\ref{fig:tps1} and Fig.~\ref{fig:tps2}. We observe that the normalized timely throughputs (timely throughput/$N$) converge to the asymptotic ($N\to\infty$) timely throughputs quite quickly. Even with the scale $N=4$, the normalized timely throughput has equilibriated to the asymptotic timely throughput.  

Secondly, we observe that the performance obtained by using the $r^\star$ that was derived in Theorem~\ref{th:csma} is near-optimal. Since the cumulative mean arrival rate for the network shown in Fig.~\ref{fig:simutopo1} is equal to $1.6$ units, its maximum achievable normalized timely-throughput is less than or equal to $1.6$ pkts/time-slot. As seen in Fig.~\ref{fig:tps1}, the timely throughput of the optimal scheme is quite close to this upper bound.

\begin{figure}[h]
	\centering
	\includegraphics[width=0.4\textwidth]{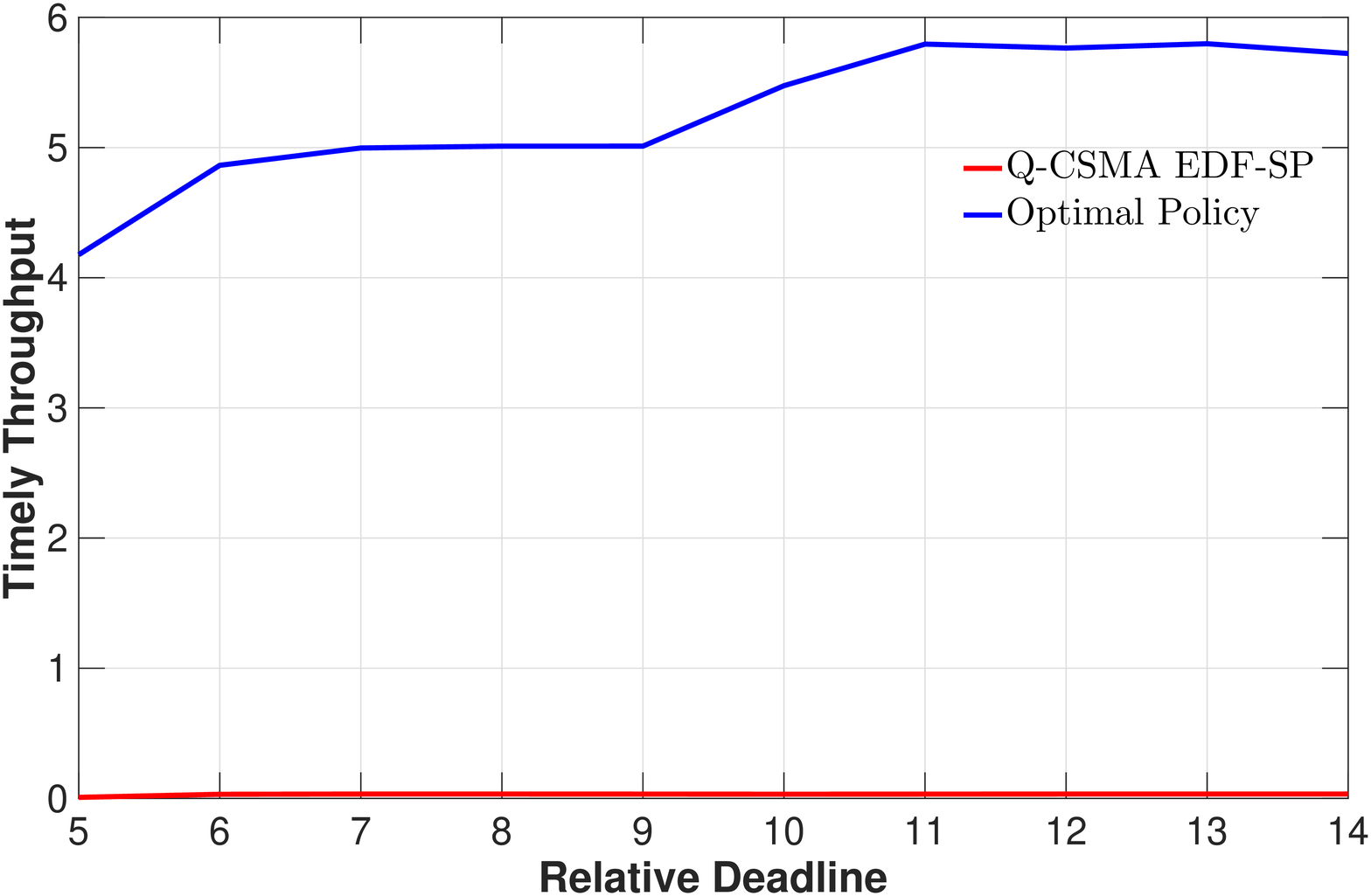}
	\caption{A plot of the timely throughput of the network shown in Fig.~\ref{fig:simutopo1} as the relative deadline of the flows is varied. Network scale, i.e., the parameter $N$ of Theorem~\ref{th:scale} and Theorem~\ref{eq:asymptot} is set to $4$. We observe that the Q-CSMA based EDF-SP policy performs poorly with respect to bandwidth allocation amongst the packets since it does not use optimally the ``information" regarding the relative deadlines and ages of the packets. In contrast, the optimal policy uses the link prices $\lambda_{\ell}$ in order to make routing-scheduling decisions, and hence attains a much higher timely throughput. The link prices enable us to prioritize the packets in proportion to the probability of their successful delivery to the destination wihin their deadline.  }
	\label{fig:tpd1}
\end{figure}

\begin{figure}[h]
	\centering
	\includegraphics[width=0.4\textwidth]{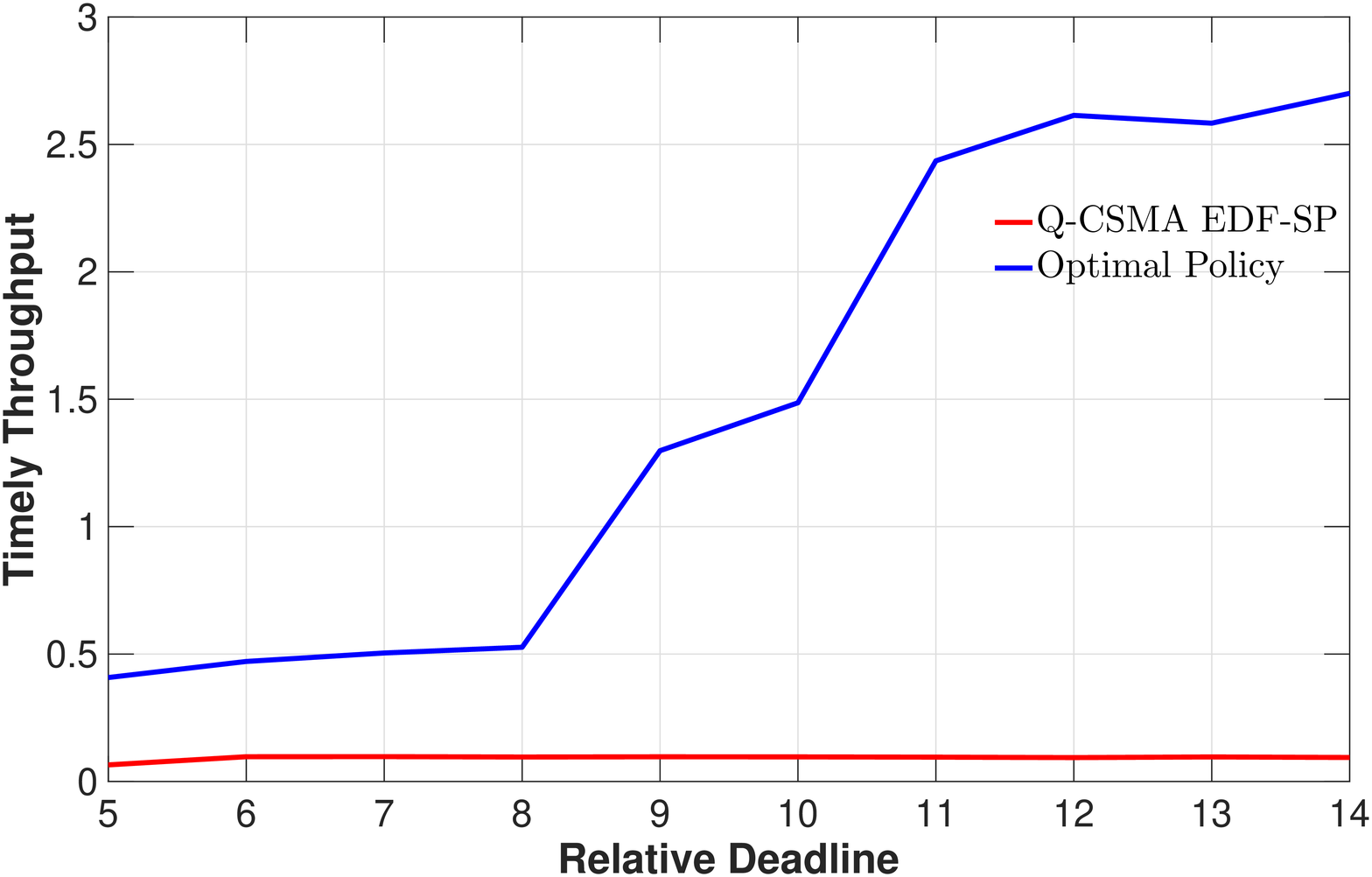}
		\caption{A plot of the timely throughput of the network shown in Fig.~\ref{fig:simutopo2} as the relative deadline of the flows is varied while keeping the network scale (the parameter $N$ of Theorem~\ref{th:scale} and Theorem~\ref{eq:asymptot}) fixed at $N=4$. Similar to the observation made in plot of Fig.~\ref{fig:tpd1}, we observe that the policy of Theorem~\ref{th:scale} outperforms the Q-CSMA based EDF-SP policy by a huge margin.}
	\label{fig:tpd2}
\end{figure}

We then fix the scale of the networks at $N=4$, and vary the end-to-end relative deadlines for the flows. The results are plotted in Fig.~\ref{fig:tpd1} and Fig.~\ref{fig:tpd2}. We observe that the performance of the optimal policy is much superior to that of the Q-CSMA with EDF-SP. This is primarily because it utilizes the link prices $\lambda_{\ell}$ in order to make decisions.  Since this automatically allows the packets to be prioritized according to the probability that they will be able to reach their destination within the deadline.  

\bibliographystyle{IEEEtran}
\bibliography{combinedbib}
\section{Appendix}
\subsection{Centralized Optimization of Bandwidth Allocation $\bar{I}$}\label{sec:bo}
We will now provide a centralized algorithm that performs optimization over the bandwidth allocation vector $\bar{I}$ that  solves~\eqref{eq:intavg1}-\eqref{eqintavg13}. Throughout this section we will assume that there is a centralized controller that knows the wireless network $G$, and its link-interference graph. Thus it knows the various independent sets $IS$ in the set $\mathcal{I}$. This asumption is relaxed in Section~\ref{sec:csma}, where CSMA protocol is utilized for decentralized channel access.
The algorithms discussed in this section are based on the commonly used optimization technique of gradient-descent-method, and the discussion is very straightforward. Define the following,
\begin{align}
f(C^{av}) := \max_{\pi: \bar{C}^\pi\leq C^{av}} \sum_f \beta_f \bar{D}_f,
\end{align}
where the inequality $\bar{C}^\pi\leq C^{av}$ between two vectors is to be taken componentwise. 
Our concern in this section will be to solve the following optimization problem,
\begin{align}
\max &\qquad f(C^{av})\label{eq:adhoc1}\\
\mbox{ s.t. } C^{av}_\ell &= C_\ell \left(\sum_{m:\ell\in C_m} \bar{I}_m \right),\forall \ell \in E,\label{eq:adhoc2}\\
\mbox{ and } \sum_{m}\bar{I}_m &= K.\label{eq:adhoc3}
\end{align}
\begin{assumption}

\end{assumption} We will design an iterative algorithm based on gradient descent that will converge to the value of $\bar{I}^\star$, which is the optimal bandwidth allocation vector. Denote by $\bar{I}^k$ the value of average bandwidth allocation vector at iteration $k$. The central controller keeps track of $\bar{I}^k$. It then updates it according to,
\begin{align}\label{eq:subgradI}
\bar{I}_m^{k+1} = \bar{I}^k + \alpha^k \frac{\partial f }{\partial \bar{I}_{m}}, m = 1,2,\ldots,M,
\end{align}
where the quantity $\frac{\partial f }{\partial \bar{I}_{m}}$ is the rate of change of the maximum achievable timely-throughput under the constraint on average bandwidth consumption on link $\ell$ set to $C_{\ell} \left(\sum_{m:\ell\in IS_m} \bar{I}_m \right)$. The iterates stated above are also projected onto the feasible set $\{\bar{I}: \bar{I}_m\geq 0~\forall m,\sum_{m}\bar{I}_m = K\}$. Next, we derive an explicit expression for the quantities $\frac{\partial f }{\partial \bar{I}_{m}}$.

\begin{lemma}\label{lemma:shadow}
For the problem~\eqref{eq:p1}-\eqref{eq:c1} since the Lagrange multiplier $\mu_\ell$ associated with the constraint $\bar{C}_{\ell}^\pi \leq C^{av}_{\ell}$ can be interpreted as shadow price, we have,
\begin{align*}
\frac{\partial f}{\partial C^{av}_\ell} = \mu^{\star}_\ell(C^{av}), \forall \ell \in \mathcal{E},
\end{align*}
where $\mu^{\star}(C^{av})$ is the vector that solves the dual problem~\eqref{eq:dualprob} and can be obtained by the gradient descent method as discussed in Section~\ref{subsec:mu}. Since the average bandwidth available to a link $\ell$ is the sum of the bandwidths provided to each independent set that it is part of, we have that 
\begin{align*}
\frac{\partial f}{\partial \bar{I}_m} = \sum_{\ell\in IS_m} \mu^{\star}_\ell(C^{av}).
\end{align*}
\end{lemma}
\subsection{Online Learning using Multiple Time-Scales Stochastic Approximation}\label{sec:3layer}
We now use the technique of stochastic approximation~\cite{robbins,kushner2012stochastic,borkarbook} in order to solve the problem of searching for the optimal policy in case the network parameters are unknown. Recall that the following 3 iterations were performed
\begin{enumerate}
\item Value Iterations for solving Dynamic Programming equations~\eqref{eq:rl}, which yield the policies $\pi^{\star}_f(\mu)$.
\item Gradient descent iterations~\eqref{eq:priceupd} for solving the dual problem~\eqref{eq:dualprob}.
\item Gradient descent iterations~\eqref{eq:csmalearn1} for obtaining the locally optimal CSMA aggression parameter $r^{\star}$ in problem~\eqref{eq:adhoc13}-\eqref{eq:csmaopt}.
\end{enumerate}
We will now combine these algorithms using multi timescale stochastic approximation~\cite{borkarscale,borkarbook} so as to obtain a single online learning algorithm.

We note that the constraints~\eqref{eq:c1} imposed in the definition of $f(C^{av})$
 involve average values of link bandwidth consumption and hence the instantaneous bandwidth utilization by the optimal policy corresponding to the evaluation of $f(C^{av})$ can exceed $C^{av}_\ell$.\footnote{Indeed, for the policy that is optimal under link-level average bandwidth constraints, if the instantaneous state of the packets present at time $t$ at a link $\ell$ is such that their cumulative bandwidth demand is in excess of $C^{av}_{\ell}$, then the link $\ell$ would simply charge them a price of $\mu^{\star}_\ell$, and provide them excess capacity.}
In contrast, the CSMA-$r$ access protocol is oblivious to the state of the packets present at various network links. It allocates bandwidth in an i.i.d. fashion, without any knowledge of the state of the packets present at various links $\ell\in\mathcal{E}$, and it also imposes a hard constraint on the cumulative bandwidth consumption, i.e. $\sum_m I_{m}(t) = N$.

Hence, we make the following crucial assumption. 
 \begin{assumption}\label{assum:excess}
 During the ``learning" phase of network operation when we are solving for the optimal CSMA parameter $r^{\star}$, we will allow the links to utilize bandwidths in excess of that provided by the CSMA scheme. This can be achieved by assuming that each link $\ell\in \mathcal{E}$ has spare bandwidth available to it during this phase, and moreover this bandwidth is reserved for its usage, i.e., no link interference takes place.
 \end{assumption}
In view of the above discussion, we will use different symbols to denote the bandwidths allocated by the CSMA and that \emph{actually} utilized by the network. 
 
\begin{figure}[h]
	\centering
	\includegraphics[width=0.4\textwidth]{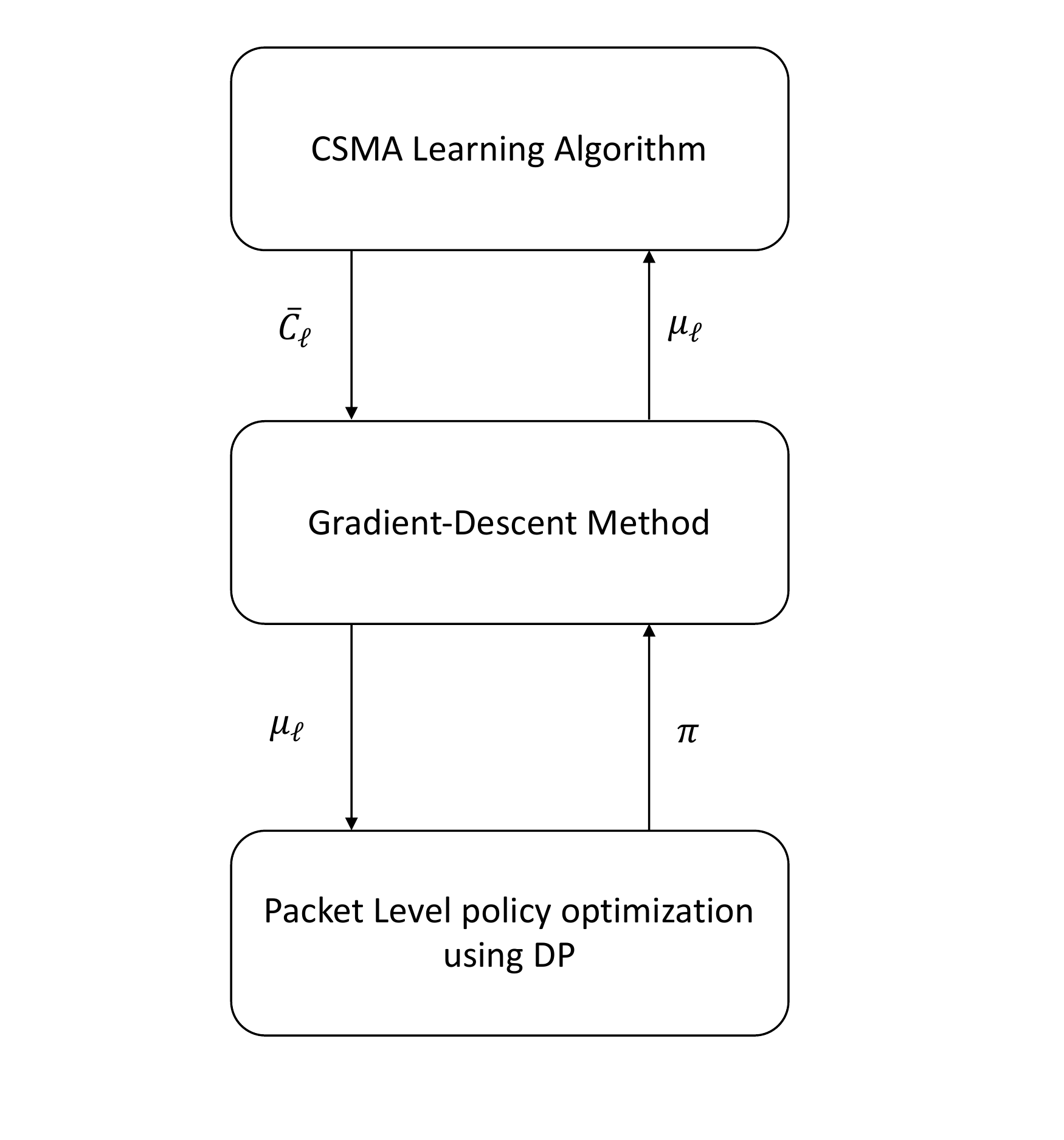}
	\caption{A hierarchial view of the 3 components of our proposed algorithm. We notice that so far we have assumed that the individual components can be computed instantaneously.}
	\label{fig:algo-decom}
\end{figure}

The bandwidth allocated by CSMA protocol, i.e., $C^{CSMA}_{\ell}(t)$, may not be equal to the actual bandwidth utilized by the routing policy, which is denoted $C^{U(t)}_{\ell}(t)$. The superscript $U(t)$ denotes that the value is decides by the actual scheduling decision implemented at time $t$. We note that typically $C^{U(t)}_{\ell}(t)$ will necessarily exceed $C^{CSMA}_{\ell}(t)$ not only because the Q-learning iterations make packet-based decisions, but also due to the fact that price iterations that solve the dual problem rely on bandwidth utilization $C^{U(t)}_{\ell}(t)$ exceeding the bandwidth availability $C^{CSMA}_{\ell}(t)$. We will thus use Assumption~\ref{assum:excess}. Furthermore since we are searching for the value of $r$ that maximizes the timely throughput under average bandwidth constraints, the scaling parameter $N$ is irrelevant, and the algorithm proposed below can be used with $N=1$.
\begin{theorem}\label{th:3layer}
We propose the following $3$-layered iterative algorithm that converges to the optimal CSMA parameter $r^{\star}$ that solves the problem~\eqref{eq:adhoc13}-\eqref{eq:csmaopt}, and the optimal policy for the relaxed problem $\pi^{\star}(C(r^{\star}))$. 
\begin{align}
&Q^f(i,\tau,j) = Q^f(i,\tau,j)\left(1-\alpha_t\right)\\
&+\alpha_t\left\{  -\mu_{(i,j)} + \mathbbm{1}(i=d_f) + \max_{\tilde{j}}Q(i^{+},(\tau+1)\wedge B,\tilde{j})\right\},\label{eq:adhoc7}\\
\tag*{Q-Learning}\\
&\mu_{\ell}(t+1) = \Gamma\left[ \mu_\ell(t) + \beta_t \left(C^{U(t)}_{\ell}(t)-C^{CSMA}_{\ell}(t) \right)   \right],\label{eq:adhoc8}\\
\tag*{Price Gradient Descent}\\
&r_\ell(t+1) \\
&= \Gamma\left(r_\ell(t) + \gamma_t \left\{ \mu_{\ell}(t) \left(C^{CSMA}_\ell(t) - C^{CSMA}_\ell(t)^2\right)        \right\}\right), \label{eq:adhoc9}\\
\tag*{CSMA optimization}\\
&t = 1,2,\ldots.\notag
\end{align}
where $\sum_t\alpha_t = \infty,\sum_t\alpha^2_t<\infty,\sum_t\beta_t = \infty,\sum_t\beta^2_t<\infty,\sum_t\gamma_t = \infty,\sum_t\gamma^2_t<\infty$ and also $\beta_t=o(\alpha_t),\gamma_t=o(\beta_t)$. 
The roles of the three layers that comprise the algorithm are described below.
\begin{enumerate}
\item Layer $1$ [Q Learning]: Learns the optimal scheduling policy to solve the \emph{Single Packet Transportation Problem} that is paramterzied by the current link prices $\mu(t)$.
\item Layer $2$ [Online Gradient Descent for Price ]: is provided ``target link capacities $\{C^{CSMA}_\ell\}_{\ell\in\mathcal{E}}$ from \textbf{Layer $3$}, and adjusts the link-prices $\{\mu_\ell\}_{\ell\in\mathcal{E}}$ so that the ``average traffic intensity" resulting from the policy produced by Layer $1$, i.e., $C^{tr}$ respects the target link capacities $C^{CSMA}$. Iterations are based on the sub-gradient descent method~\eqref{eq:priceupd}.
\item Layer $3$ [CSMA Learning] : gets access to the link prices $\{\mu_\ell\}_{\ell\in\mathcal{E}}$ from Layer $2$, and modulates the aggression parameter $r$ of the CSMA-$r$ protocol. It performs the bandwidth optimization by converging to the optimal aggression rate $r^{\star}$.
\end{enumerate}
\end{theorem}
Figure~\ref{fig:algo-decom} depicts the 3-layered hierarchial structure of the proposed algorithm.
\begin{remark}
Notice that the iterations asociated with tuning the parameter $r$ can be guaranteed to converge to only a local optima of the function $F(\cdot)$. However, gradient descent schemes in general suffer from this drawback unless the function $F(\cdot)$ is shown to be convex. For our problem, it is not easy to the convexity of the function $F(r)$ in the parameter $r$. One can use noisy perturbations (see~\cite{kushner,kushner2012stochastic}) in order to ensure that the iterations do not get stuck at a critical point that is not local optima. If we utilize methods like simulated annealing, then the $r$ iterations are guaranteed to converge to a global optima. However the convergence speed of simulated annealing procedure is too slow to be of any practical application. 
\end{remark}
We realize that due to the $3$ layered structure, and complex multi-timescale nature of the above algorithm, the proposed algorithms may not be practical to implement. Thus, we now consider a somewhat related problem that involves scheduling packet transmissions under constraints on link-bandwidths. Next, we introduce this objective. 

\end{document}